\documentclass[12pt]{jpsj2}

\usepackage[dvips]{color}

\title{Analyzing the House Fly's Exploratory Behavior with Autoregression Methods}
\author{Hisanao TAKAHASHI$^{1}$\thanks{Previously, Graduate School of Arts and Science, University of Tokyo. E-mail: takahashi@sp.dis.titech.ac.jp }, Naoto HORIBE$^{2}$, Masakazu SHIMADA$^{2}$,\\
Takashi IKEGAMI$^{2}$
}
\inst{
$^{1}$Interdisciplinary Graduate School of Science and Engineering,\\
Tokyo Institute of Technology. Yokohama 226-8502\\
$^{2}$Graduate School of Arts and Science, University of Tokyo. Tokyo 153-8902}

\date{}

\kword{
housefly, AR model, anomalous diffusion, exploration behavior, memory and learning, autonomous dynamics
}

\abst{
This paper presents a detailed characterization of the trajectory of a single housefly with free range of a square cage. The trajectory of the fly was recorded and transformed into a time series, which was fully analyzed using an autoregressive model, which describes a stationary time series by a linear regression of prior state values with the white noise. 
The main discovery was that the fly switched styles of motion from a low dimensional regular pattern to a higher dimensional disordered pattern. This discovered exploratory behavior is, irrespective of the presence of food, characterized by anomalous diffusion.
}

\begin{document}
\maketitle

\section{Introduction}
Biological autonomy is an outstanding feature of living systems, one which even the most complex man-made robots fail to show. Different from the mathematical definition of autonomy, biological autonomy does not simply mean decoupling from an environment; instead, it is enhanced by coupling with an environment. To understand the nature of biological autonomy, we designate the common fly as a test subject to investigate its autonomous behavior.
 
The advantage of studying a fly's behavior is that a fly has a variety of navigational patterns and responds to environmental information. Their responses are not simply reactive, as they may behave differently even in the same context. A fly's navigational pattern can not be simply described as a Brownian motion curve but may be more complex function of the coupling a fly and its environment.

 From a theoretical point of view, we define and study autonomous motion as an interplay between internal and external dynamics, which was recently proposed as a variety of embodied chaotic itinerancy (ECI)\cite{ikegami}.
Chaotic Itinerancy is a relatively common feature among high-dimensional chaotic systems, which shows itinerant behaviour among low-dimensional local attractors through higher-dimensional chaos\cite{ikeda, kanekotsuda}.

Of course, we cannot directly see this interplay, but here we assume that the interplay of dynamics can be quantified by the navigational trails of a fly. In particular, this switching navigational style is what ECI considers to be a candidate of autonomous motion. A more in-depth characterization of this switching behavior is the main purpose of this paper.

Until recently, spontaneous movements like a fly's exploratory movements have not been studied quantitatively, but one of the first serious observations was provided by Murdie and Hassell\cite{Murdie+Hassell} in 1973. They analyzed the apparent changes of the fly's walking between pre- and post-feeding phases by decomposing the fly's temporal motion into turning angle and  distance (e.g. forwarding) components. Before feeding, there is considerable variation in both the angle turned and the distance moved. After feeding, there is a sharp increase in the mean angle turned and a sharp decrease in the distance moved. They also mention that a fly's biased motion increases the probability of discovery of food sites. 

We, for the first time, use an autoregression model (AR model)\cite{Kitagawa,Box+Jenkins} to analyze the trajectory data. The AR model is a linear filtering of experimental data with white noise, which tells us different aspects of the time series patterns from the previous studies,\cite{Murdie+Hassell,Martin(2004)}  where their method is often called descriptive statistics. Although the AR model is one of the simplest models that examines time structure, it is nevertheless effective in classifying the dynamics, as we show in this paper.

In our analysis of a local stationary AR model, we found that the distance movement followed a stationary distribution, which is rather inconsistent with the results of Murdie and Hassell. We assume that the difference comes from the definition of ¡Èstationary¡É in our experiment. Our definition is appropriate if we regard a fly as an autonomous agent.  Additionally, when a fly changes its behavioral pattern, this change must be correlated with internal dynamics as organized by the memory and genetics. 
Also, by analyzing the difference between the fly's navigation and a random walk, which has been discussed in terms of the foraging efficiency,\cite{Bond,Klafter(2005), Viswanathan(1999), Fernandez(2004),Takahashi(2003)} we will also show that a fly's navigation pattern shows an anomalous diffusion process irrespective of the existence of food.

\section{Method}

\subsection{Experiment}

\begin{figure}[htbp]
\hspace{2cm} CCD camera\\
\includegraphics[height=5cm,]{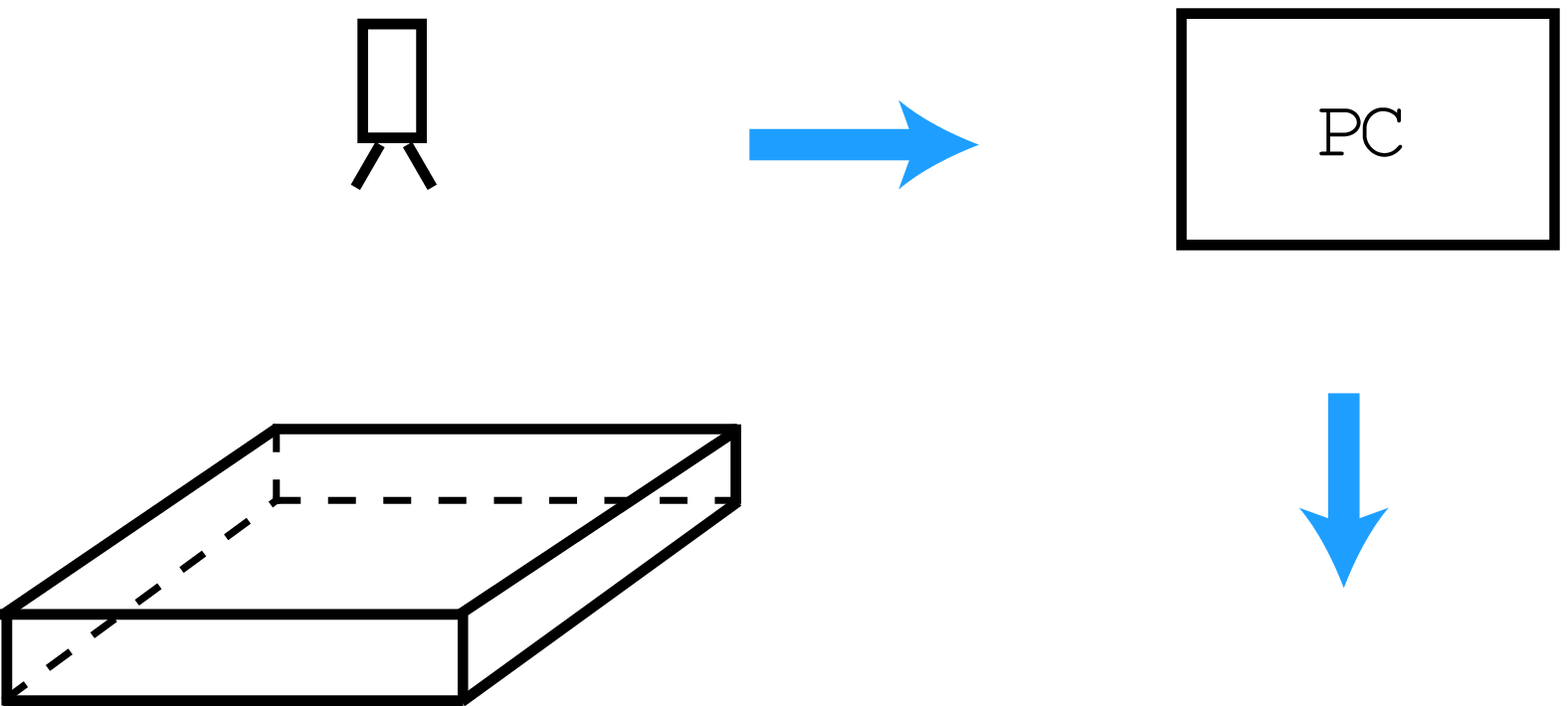}\\
~~~~~~~~ Acryl cage ~~~~~~~~~~~~~~~~~~~~~~~~~~~~~~~~~~~~~~~~~~~ Time series\\
~~~~~~~~ $ 47\times 47 \times 2.5$ cm\\
\caption[]{Experimental system}
\label{fig. System of experiment}
\end{figure}

\begin{figure}[htbp]
\begin{center}
	\rotatebox{90}{\hspace{5cm}\rotatebox{-90}{$  $} }
	\includegraphics[height=10cm]{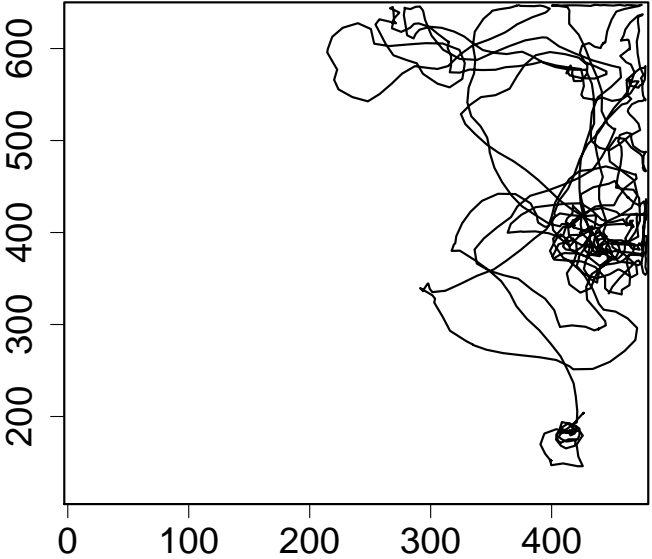}
\end{center}
\caption[]{Trajectory of a fly for about 7 minutes with sugar solution's droplets.}
\label{fig.trajectory 051207all}
\end{figure}

We used a common housefly and a personal computer to record the trajectory of the fly's walking behaviors in an acrylic cage by digital video camera. Figure \ref{fig. System of experiment}  shows the overall setup of our system. To 
begin with, a fly is put in the acrylic cage whose size is 47 cm square and 2.5 cm deep. From the cage above, the digital video camera takes a picture of the whole layout. Then, the place of the fly in the picture is transformed into a two-dimensional time series (in x, y coordinates) using a personal computer. 
Small droplets of sugar solution (4 \% source, 2 ml) were distributed on the floor of the cage. 
We also performed some experiments under the no-sugar solution condition for comparison. 
Figure \ref{fig.trajectory 051207all} shows one trajectory of a fly for about 7 minutes.

\subsection{Local stationary AR model}

We use a local stationary autoregressive (AR) model which only considers the local time series by the following procedure.
Let us denote the value of time series at time $ t,\ t-1,\ t-2,... $ by $x_t,\ x_{t-1},\ x_{t-2}...$.  Also let $ z_t,\ z_{t-1},\ z_{t-2},...$ be the deviations from the mean value of the time series $ \mu $, i.e. $ z_t = x_t-\mu $.
 Then, the $ m$th order AR model is defined as the dimension of the linearly spanned space;
\begin{eqnarray}
	z_t = a_1\, z_{t-1}+ a_2\, z_{t-2}+\cdots+a_m\, z_{t-m} +w_t,
\end{eqnarray}
where $ w_t $ is the Gaussian white noise whose  variance is $ \sigma^2 $.
A standard Akaike Information Criteria (AIC)\cite{Akaike,Kitagawa} is used to decide the values of the AR coefficients $ a_1,\ a_2,...,\ a_m$, the variance $\sigma^2$ and the AR order $ m $. The number of the AR order $ m $ is called the AR dimension.

\begin{figure}[htbp]
\hspace{1.3cm}${\rm AIC}_1$\hspace{.7cm}${\rm AIC}_2$\\
\hspace{.6cm}\includegraphics[height=3cm]{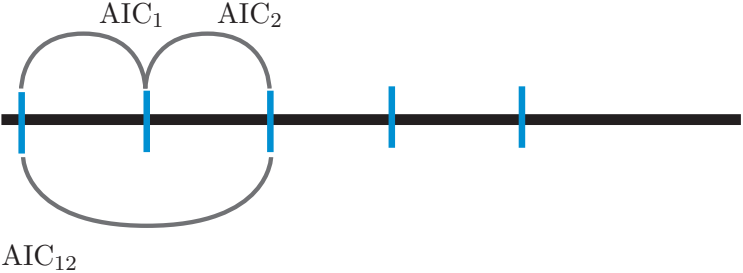}
\vspace{-5mm}\\
\hspace{2cm}${\rm AIC}_{12}$\\
\caption[]{Local stationary AR model}
\label{fig. local stationary AR model}
\end{figure}

We construct a local stationary AR model recursively by the following procedures:
\begin{description}

\item{Step 1)} Divide the time series into  well-defined small intervals which have the same segmental length $ L $, and suppose the AR model is stationary in each interval.

\item{Step 2)}
Using the AIC, we decide the AR coefficient and the variance in the first and second intervals and call these AIC values $\rm AIC_1$ and $ \rm AIC_2$, respectively 
(see Fig. \ref{fig. local stationary AR model}).

\item{Step 3)}
We make one united interval from the two successive intervals, taking the starting point of the first interval as the starting point of the united one and the ending point of the second interval as the ending point of the united one.
 We decide the AR coefficient and the variance for this interval in the same way as before, and name this AIC value as $ \rm AIC_{12} $.
If the inequality $ \rm AIC_1 + AIC_2 < AIC_{12} $ holds, then we assume the two intervals are driven by a different AR mode; otherwise, the two intervals are driven by the same AR model.
In the former case, we think the two intervals are separated segments, which 
means the walking pattern was changed between the two intervals. In the latter case, the two intervals are treated as one segment, which means the walking pattern was the same in the two intervals. 

\item{Step 4)} We rename the intervals as follows. 
When the inequality $ \rm AIC_1 + AIC_2 < AIC_{12} $ holds, we regard the second interval as a first interval and the third interval as a second interval, 
and repeat the same procedure as before from the step 2.
When the inequality is invalid, we regard the united interval as a first interval, and the third interval as a second interval and repeat the same procedure as before from the step 2. 

\item{Step 5)}We repeat the same procedure until we have performed this operation on the entire data set.

\end{description}

If we find a united interval, we say that a sequence is stationary in the interval.
The size of the segmental length $ L $ at the step 1 is 100 in this paper.
 Practically, the number of free parameters estimated by AIC should be less than $ 2 \sqrt{L} $\cite{Kitagawa}. Consequently, $ L = 100 $ is a sufficient length in this study.

\subsection{Spectrum}\label{Spectrum}

We should be careful about interpreting the local AR model. Even if a time series is separated by different local AR models, we cannot be certain that the strategy of a fly's walking is changed at that separation point. The walking pattern is also affected by the nonstationary or nonlinear effects of the fly's walk, which are not considered by the AR model. To improve the situation, we use the AR spectrum.

We inversely generate the time series to compute the power spectrum of the reconstructed time series. That is, by generating the time series of the AR model such as, 
\begin{eqnarray*}
	z_n=\sum_{i=1}^m a_i\, z_{n-i}+w_n,
\end{eqnarray*}
we then compute its power spectrum as follows.
\begin{eqnarray*}
	p(f) &=& \sum_{k=-\infty}^\infty E(z_n z_{n-k}) e^{2\pi i k f}\\
	& = &
	\frac{\sigma^2}{\displaystyle\left|1-\sum_{j=1}^m a_j e^{2\pi i j f}\right|^2}.
\end{eqnarray*}
Where E[xy] is the temporal average of two variables x and y.

\subsection{Analysis}

\begin{figure}[htbp]
(a)\vspace{-2cm}\\
	\hspace{-1cm}\includegraphics[height=10cm]{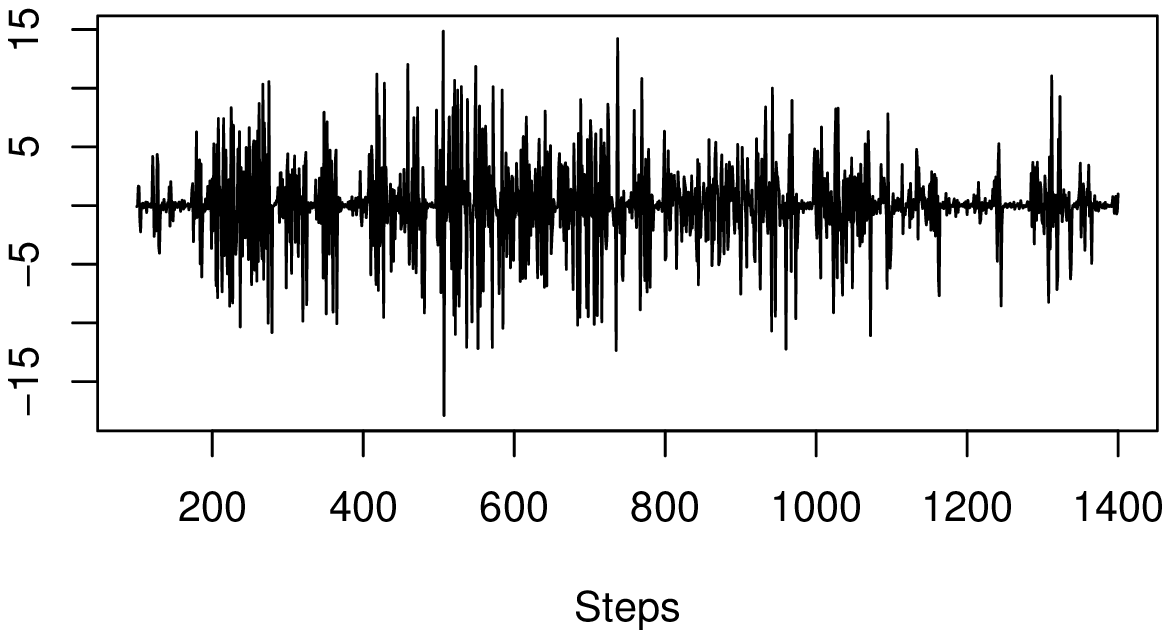}\\
(b)\vspace{-2cm}\\
	\hspace{-1cm}\includegraphics[height=10cm]{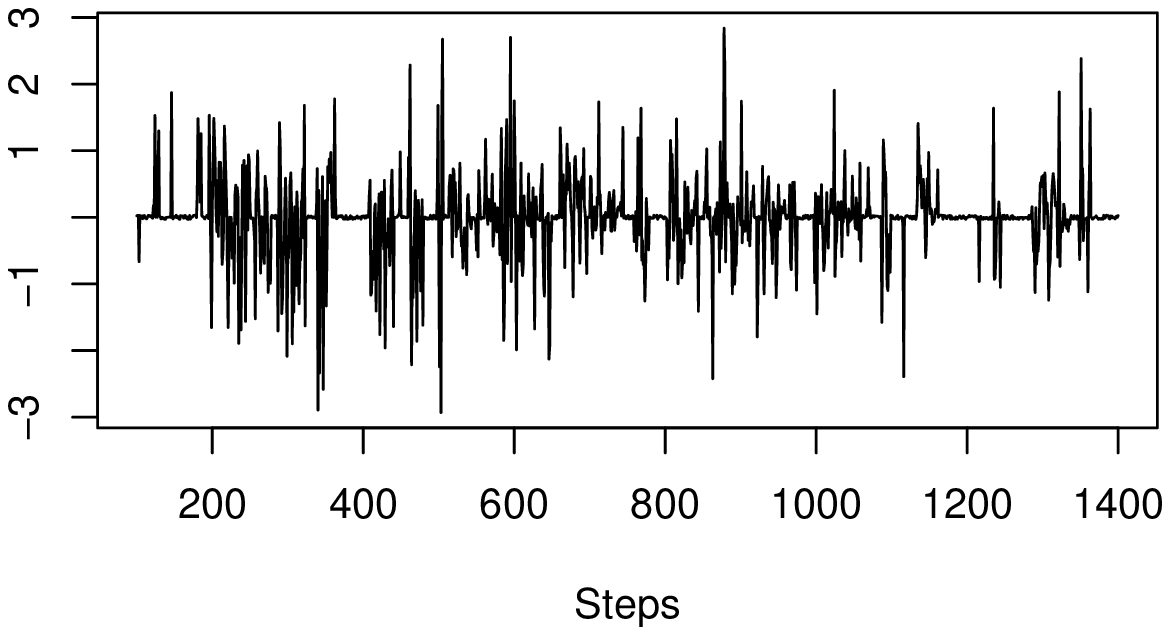}
\caption[]{Element of the velocity's difference (a) and the angular difference (b).}
\label{fig.oltho klino 100-1400}
\end{figure}

We decompose the overall motion structure into velocity $ v_i $ and angular $ \theta_i $ element components.  Before applying the local stationary AR model, we convert the target data into velocity and angular differences such as $ x_i = y_i- y_{i-1} $. Figure \ref{fig.oltho klino 100-1400}  shows the time series of the velocity difference (a) and the angular difference (b), which we will analyze practically in this paper.

\subsection{Deviation from the Brownian motion}

In order to characterize the difference between a fly's navigation pattern and Brownian stochastic motion, we study the two point correlations of the fly's navigational trajectory. Of particular interest is the ordered structure hidden in a fly's behavioral pattern which is quantified as an anomalous diffusion. Anomalous diffusion is observed in many kinds of exploratory 
behavior in organisms.\cite{Viswanathan(1999), Fernandez(2004)}  Here, we examine the diffusive speed of the fly's walk as 
\begin{eqnarray}
	\label{eq. anomalous diffusion}
	E_{t_0}[(x(t+t_0)-x(t_0))^2] \sim t^{\alpha},
\end{eqnarray}
where $ E_{t_0} $ represents the average with respect to the variable $ t_0 $ and $ x(t) $ is the site of a fly at time $ t $.
If the $\alpha$ is greater than one, it implies an anomalous diffusion.
 $\alpha = 1$ implies the Gaussian random walk.

The L\'evy flight might be the simplest explanation for this behavior. However, the appearance of the trajectory is different from a simulated L\'evy flight. The fly's exploratory pattern is relatively smooth, whereas the L\'evy walks studied in [\citen{Viswanathan(1999), Fernandez(2004)}] are a combination of straight lines. We will show that the fly's walk can be decomposed into several walking patterns which are classified by the AR model in this paper. Since it is possible to make an anomalous diffusion from a combination of other walking styles \cite{Takahashi(2003)}, anomalous diffusion is the expected result.

\section{Results}

A trajectory as shown in Fig.\ref{fig.trajectory 051207all}  is recorded for about 7 minutes and the minimum time span of the data, i.e. the span between $t$ and $t + 1$, is set at 0.2 seconds. The density of the line of the trajectory is higher at some region and lower at others. There are some droplets of sugar solution around the coordinate (440, 380) in this figure, where the lines are crowded.

\subsection{Behavior analysis}
A fly in the cage reaches a sugar solution by walking, then ingests the solution before leaving. After a fly leaves the solution, it walks around the solution for about one minute (from 171 steps to 470 steps), where the line becomes crowded (Fig. \ref{fig.trajectory one minute} (a)). Also, it seems that the fly's walking pattern becomes circular, almost as if it has a virtual center. We assume that this biased motion is caused by the memory capability of flies, and also that it should be consistent with the anomalous diffusion discussed later. 

Figure \ref{fig.trajectory one minute}  (b) follows the pattern of Fig. \ref{fig.trajectory one minute}  (a). The fly's walking pattern changes with its navigational style; that is, as the density of the trajectory becomes lower, the fly walks into a wider area. After searching near the previous food area for a while, the fly seems to start a search in a wider area.

\subsection{AR model}

\begin{figure}[htbp]
\begin{center}
	\rotatebox{90}{\hspace{4cm}\rotatebox{-90}{(a)} }
	\includegraphics[height=5cm]{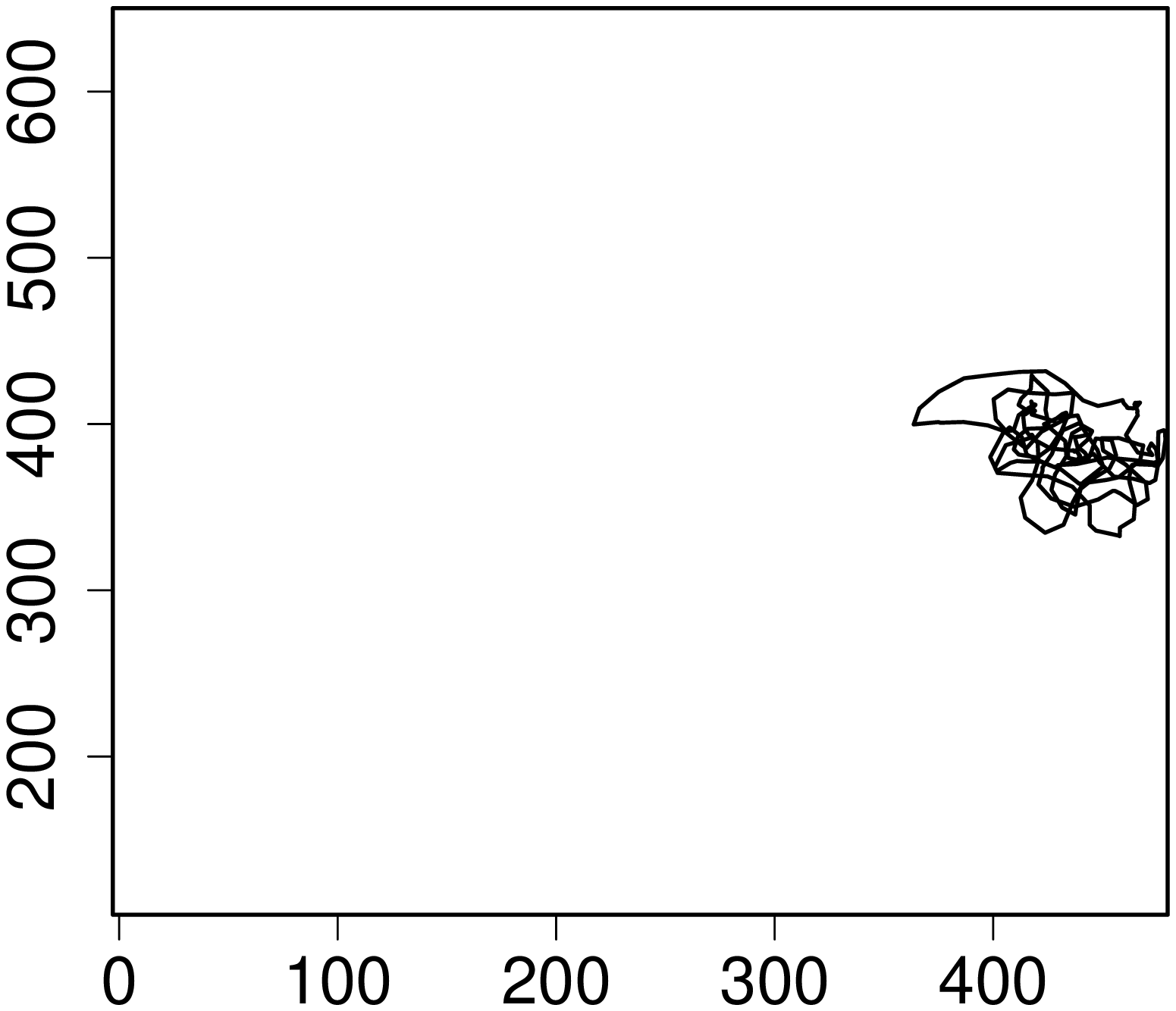}
	\rotatebox{90}{\hspace{4cm}\rotatebox{-90}{(b)} }
	\includegraphics[height=5cm]{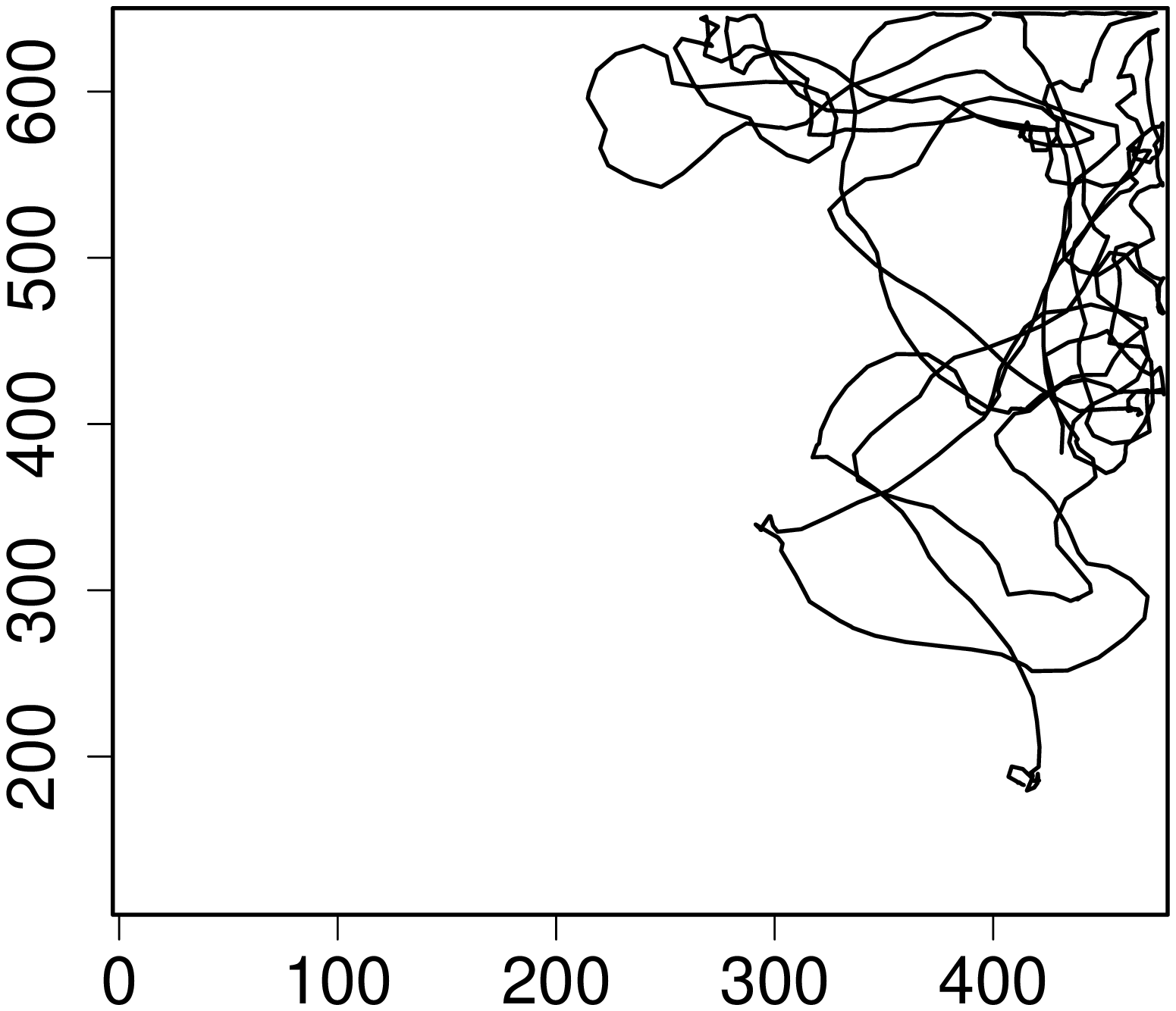}
\end{center}
\caption[]{Trajectory of a fly for one minute after feeding (a), and for about three minutes after the one minute (b). (a) is the interval from 171 to 470 steps and (b) is from 471 to 1370.
}
\label{fig.trajectory one minute}
\end{figure}

The results of the local AR model corresponding to Fig.\ref{fig.trajectory one minute} are given in Table \ref{table local AR oltho 07 L=100}.

Let us summarize the  velocity and angular data structure:

\begin{itemize}
\item[ i)] From the velocity data, two intervals from 171 steps to 370 steps are united as one interval by the local AR model with AR order 4, and from 871 steps to 1070 steps are also united with AR order 5. 

\item[ii)] From  the angular component time series, three intervals are picked  from 371 steps to 670 steps, and two from 671 steps to 870 steps,  the same AR model.

\item[iii)] Because the angular and velocity motion of the fly do not change synchronously in the cases of 271, 471, 571, 771, and 971 steps, we assume that a fly can sometimes independently control its angular and velocity motion.

\end{itemize}

As we expect from  Fig. \ref{fig.trajectory one minute}, the style of motion temporally changes.  The transition period estimated from the local AR model of the velocity data is around 371-470 but from the angular data, it is estimated around 670. This discrepancy may be due to the fact that the velocity 
and the angular element can be treated independently; this point will be discussed further later in this paper.

 In the intervals from 171 to 670, the AR order is higher with respect to the angular difference. The higher AR order represents the fact that the fly is using longer-term memory to organize the walking pattern, which is likely necessary to move around a certain point. The fly's walking in the interval from 171 to 470 corresponds to moving around the sugar solution it found (see 
Fig. \ref{fig.trajectory one minute} (a)). While it is true that we have to be concerned with both angular motion and velocity differences to explain 
the temporal behavior exhibited here, if examined under a restriction such as a constant velocity, then the winding walk is necessarily of a higher AR order.

In the intervals from 671 to 1170, and from 1271 to 1370, for the angular differences, the AR order becomes lower and the elements of the longer period are higher in its spectrum, as we will mention later, and this fact corresponds to the fact that the fly's walk is becoming smoother (see Fig. \ref{fig.trajectory one minute} (b)). In the case of the interval from 371 to 670, though the AR orders for the angular differences are high, the trajectory of the interval is smooth to some extent, so we have to consider the 
effect not only of the angular difference, but also of the velocity difference. 

\subsection{AR spectrum analysis}

\begin{figure}[hptb]
	\rotatebox{90}{\hspace{3.6cm}\rotatebox{-90}{(171-270)}}\hspace{-2.0cm}
	\includegraphics[height=4cm]{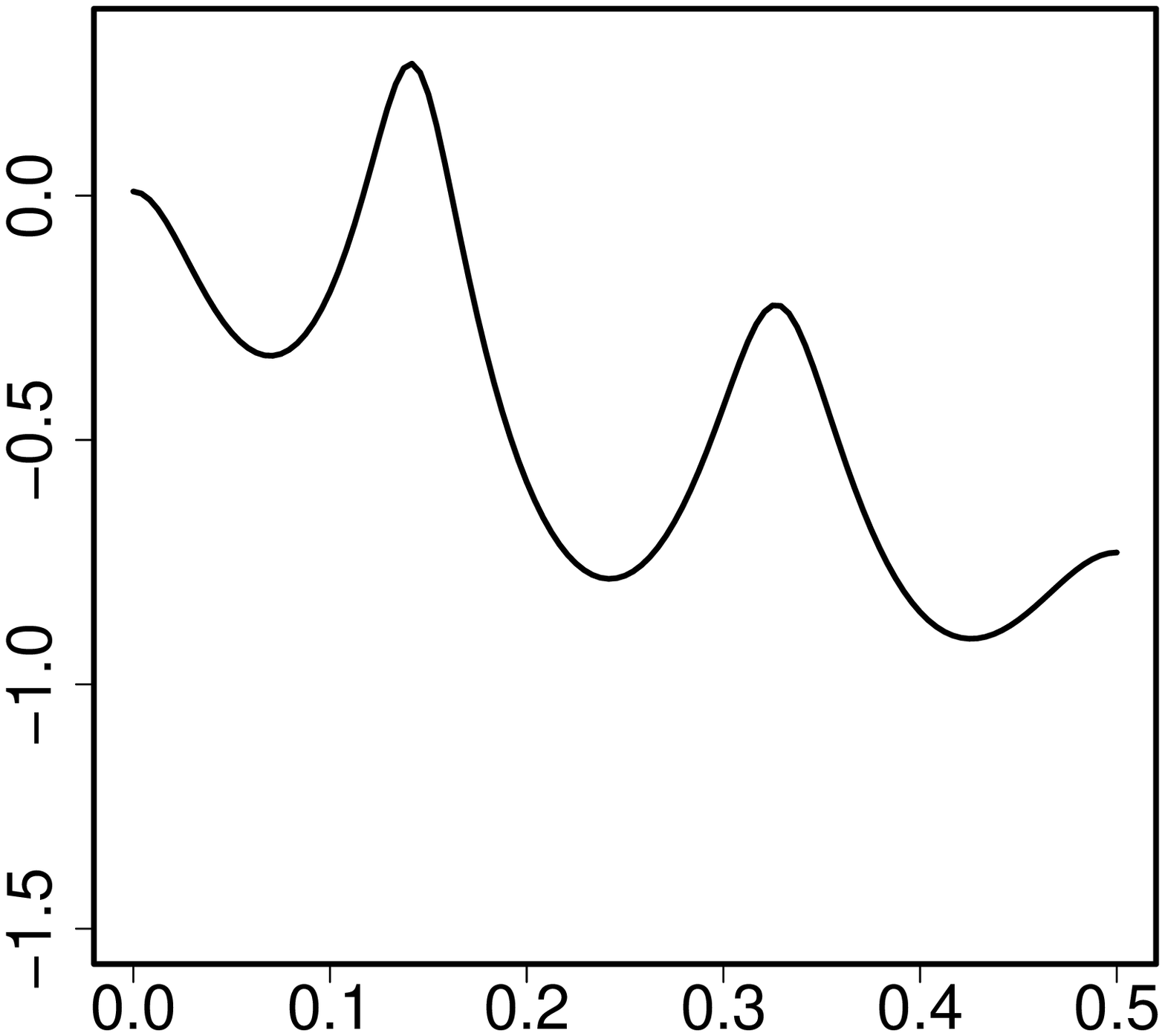}
	\rotatebox{90}{\hspace{3.6cm}\rotatebox{-90}{(271-370)}}\hspace{-2.0cm}
	\includegraphics[height=4cm]{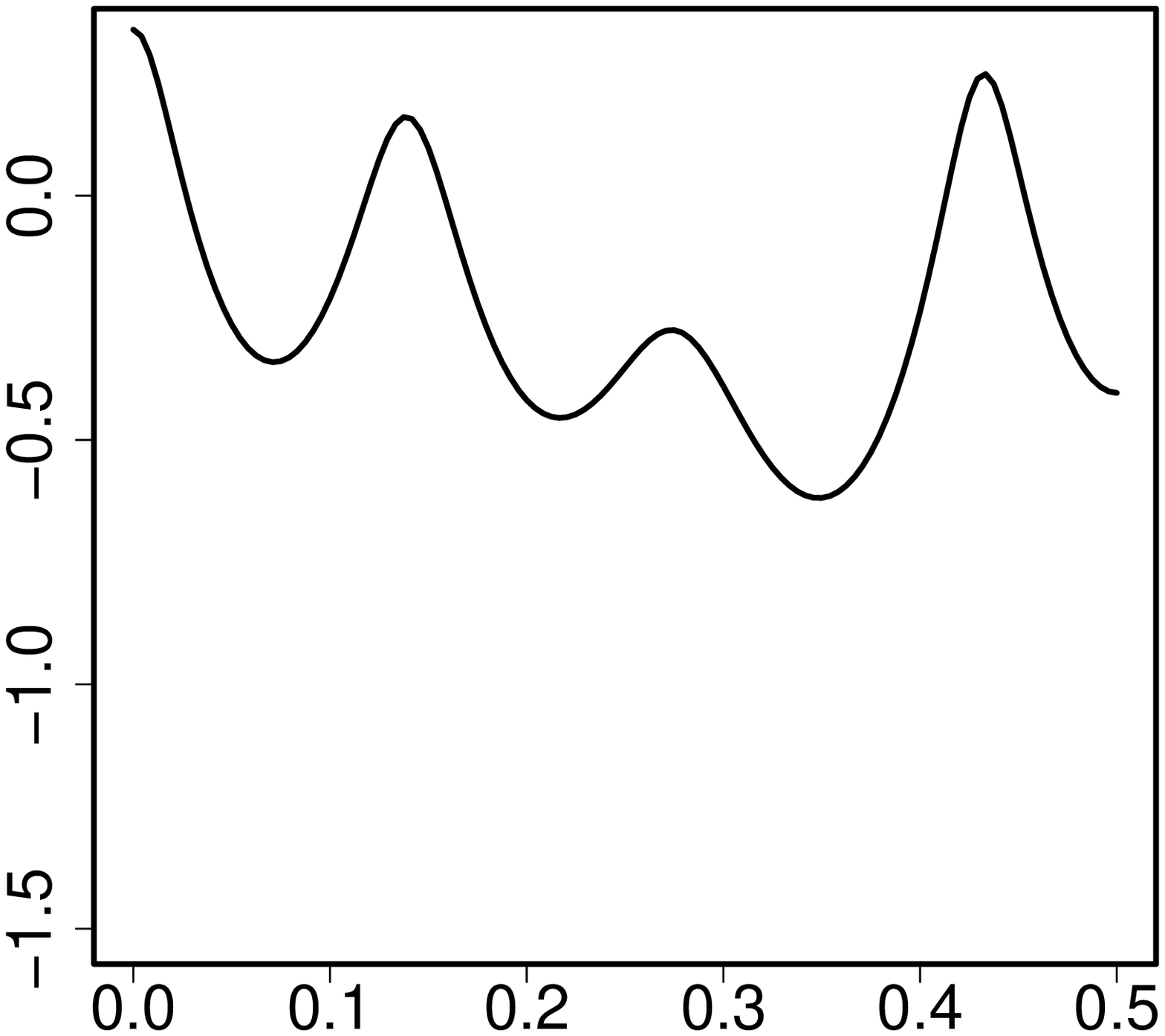}\\
	\hspace{3.7cm}
	\rotatebox{90}{\hspace{3.6cm}\rotatebox{-90}{(371-670)}}\hspace{-2.0cm}
	\includegraphics[height=4cm]{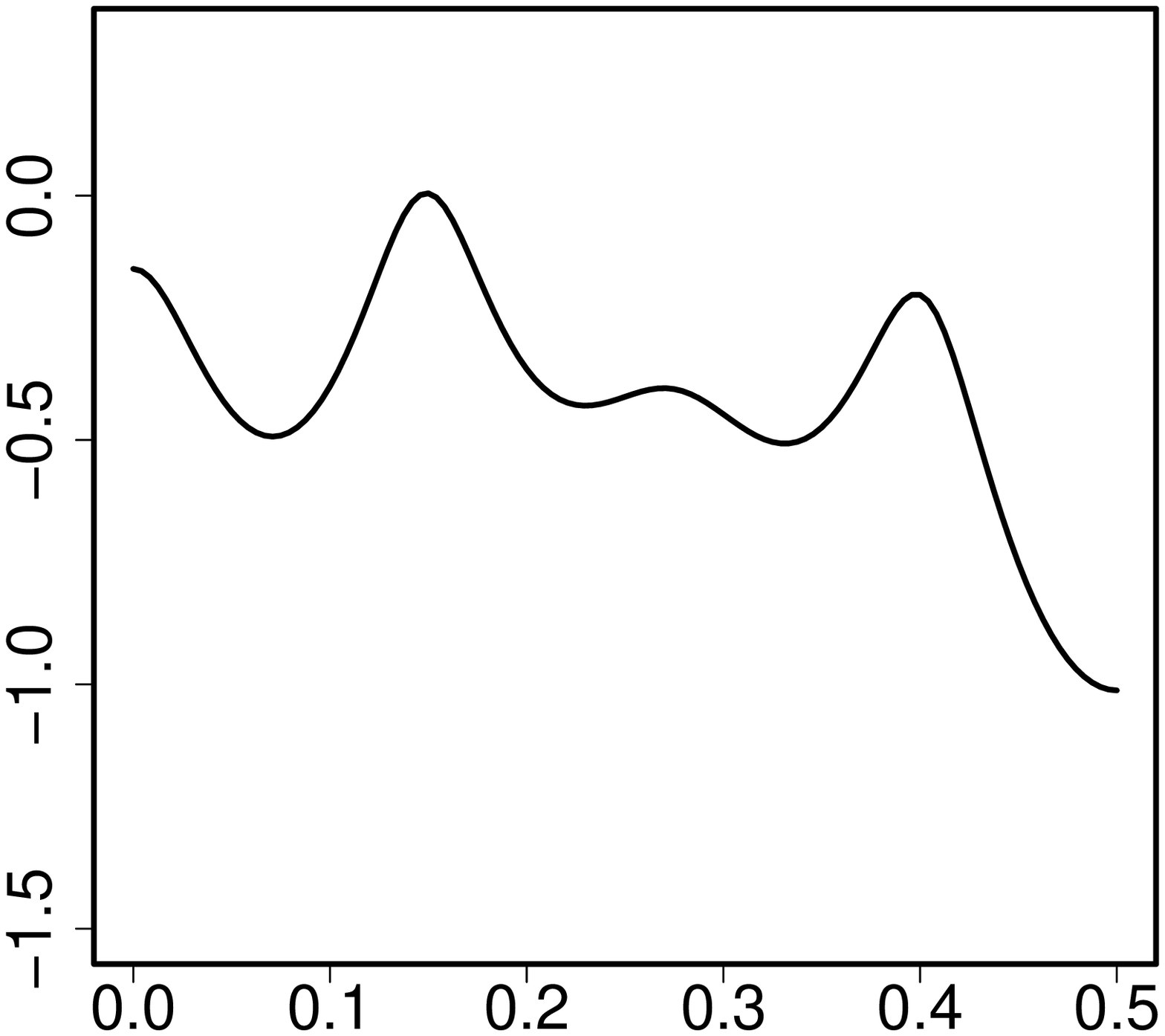}\\
	\rotatebox{90}{\hspace{3.6cm}\rotatebox{-90}{(671-870)}}\hspace{-2.0cm}
	\includegraphics[height=4cm]{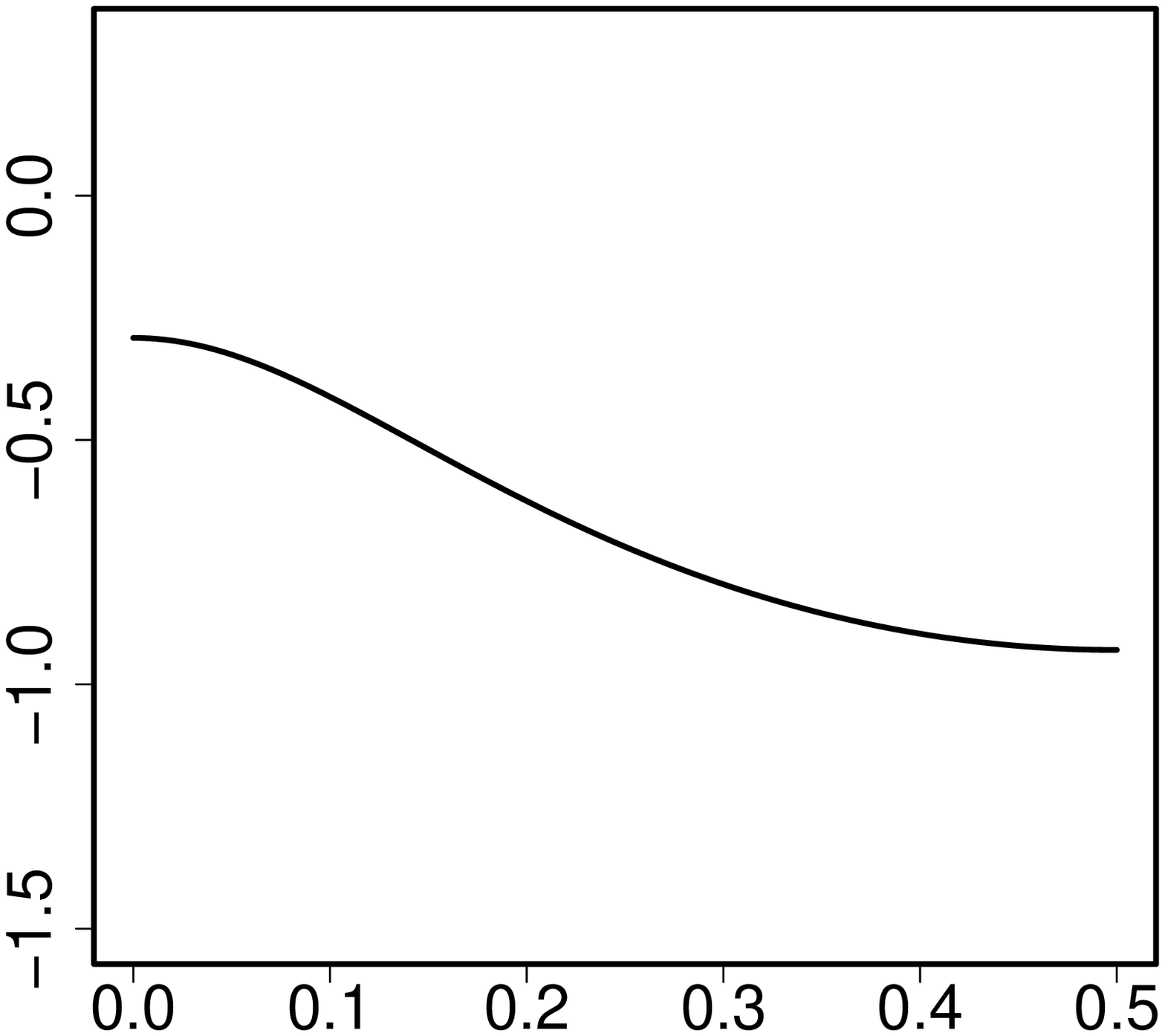}
	\rotatebox{90}{\hspace{3.6cm}\rotatebox{-90}{(871-970)}}\hspace{-2.0cm}
	\includegraphics[height=4cm]{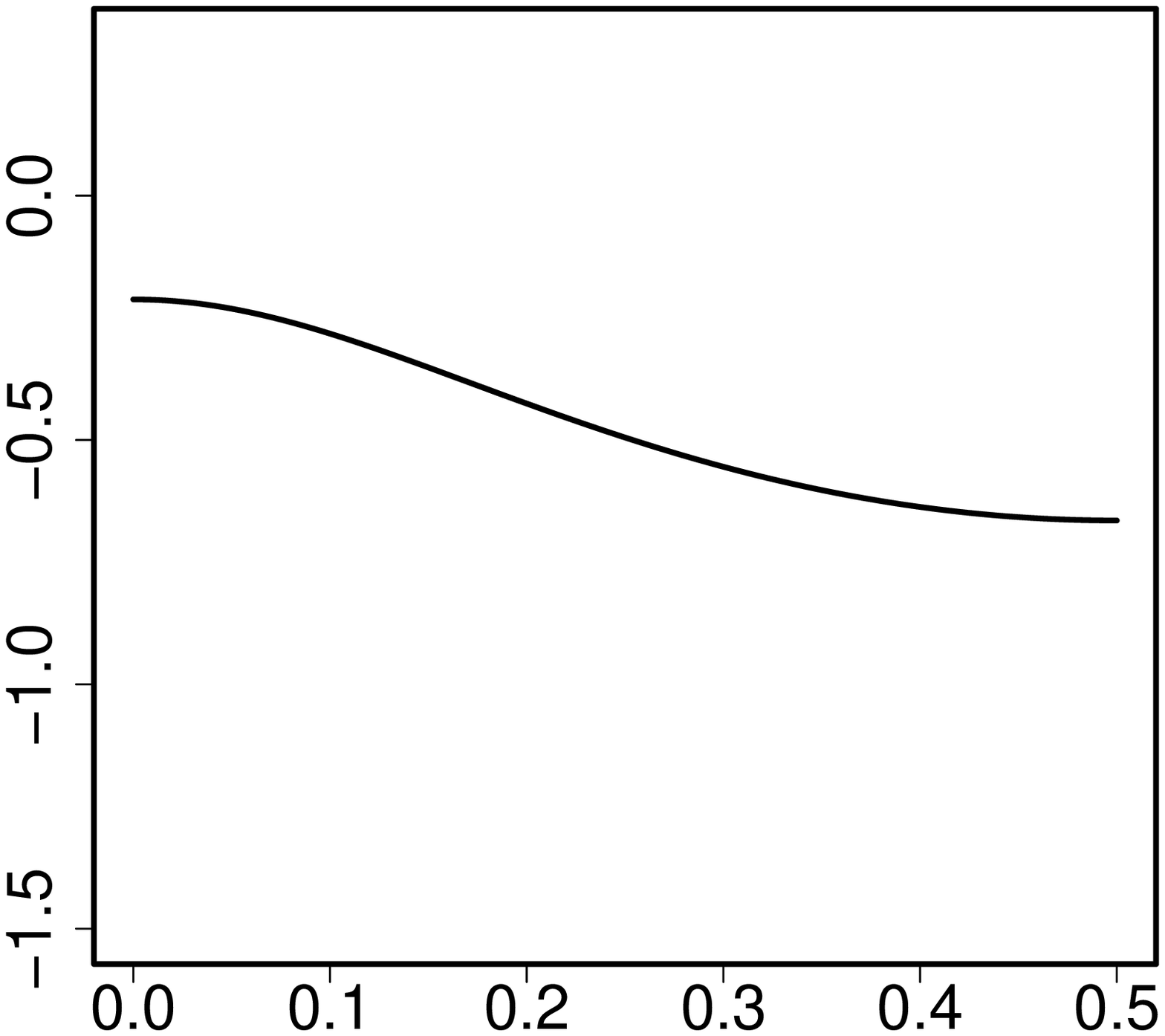}
	\rotatebox{90}{\hspace{3.6cm}\rotatebox{-90}{(971-1070)}}\hspace{-2.2cm}
	\includegraphics[height=4cm]{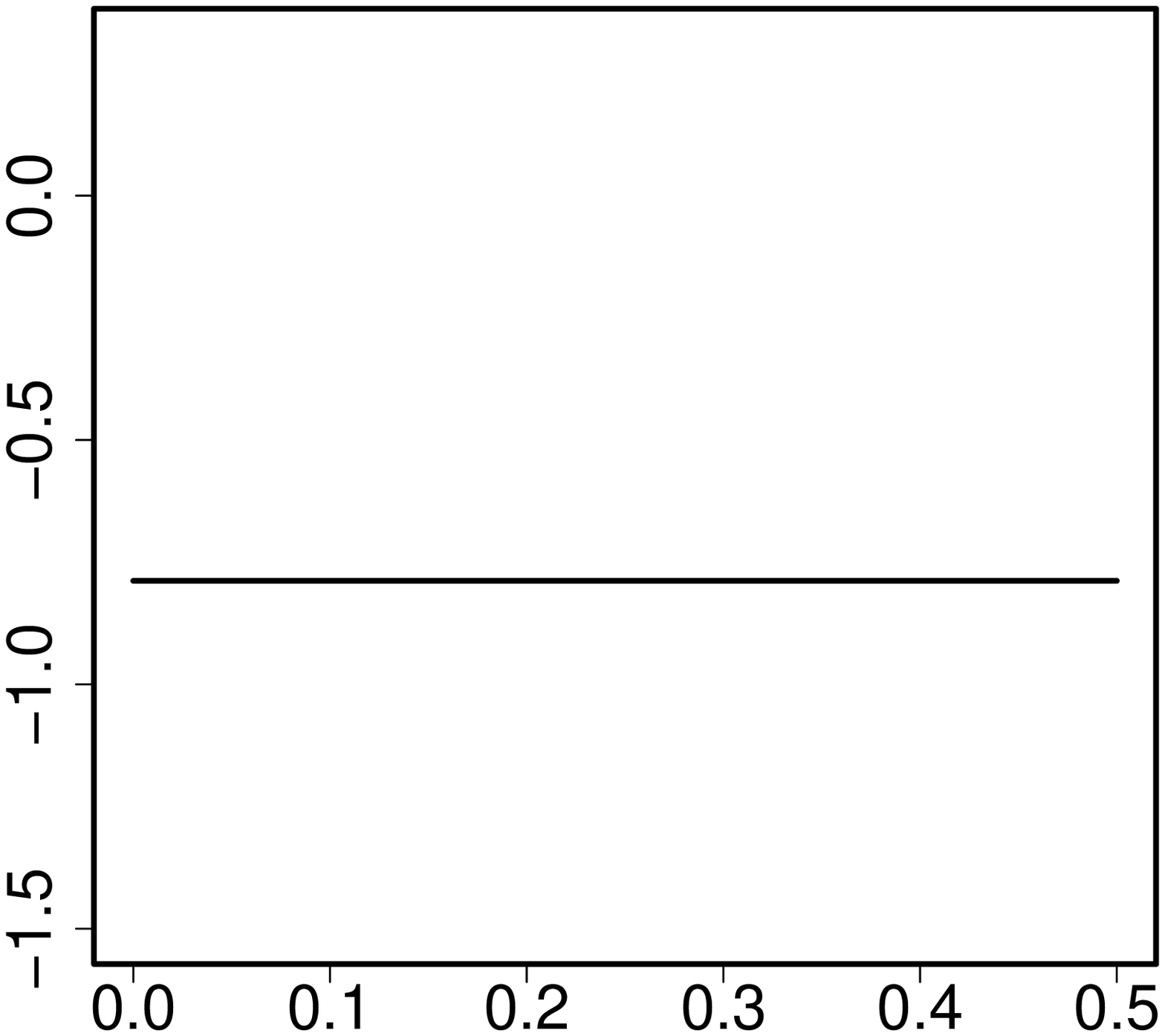}\\
	\rotatebox{90}{\hspace{3.6cm}\rotatebox{-90}{(1071-1170)}}\hspace{-2.4cm}
	\includegraphics[height=4cm]{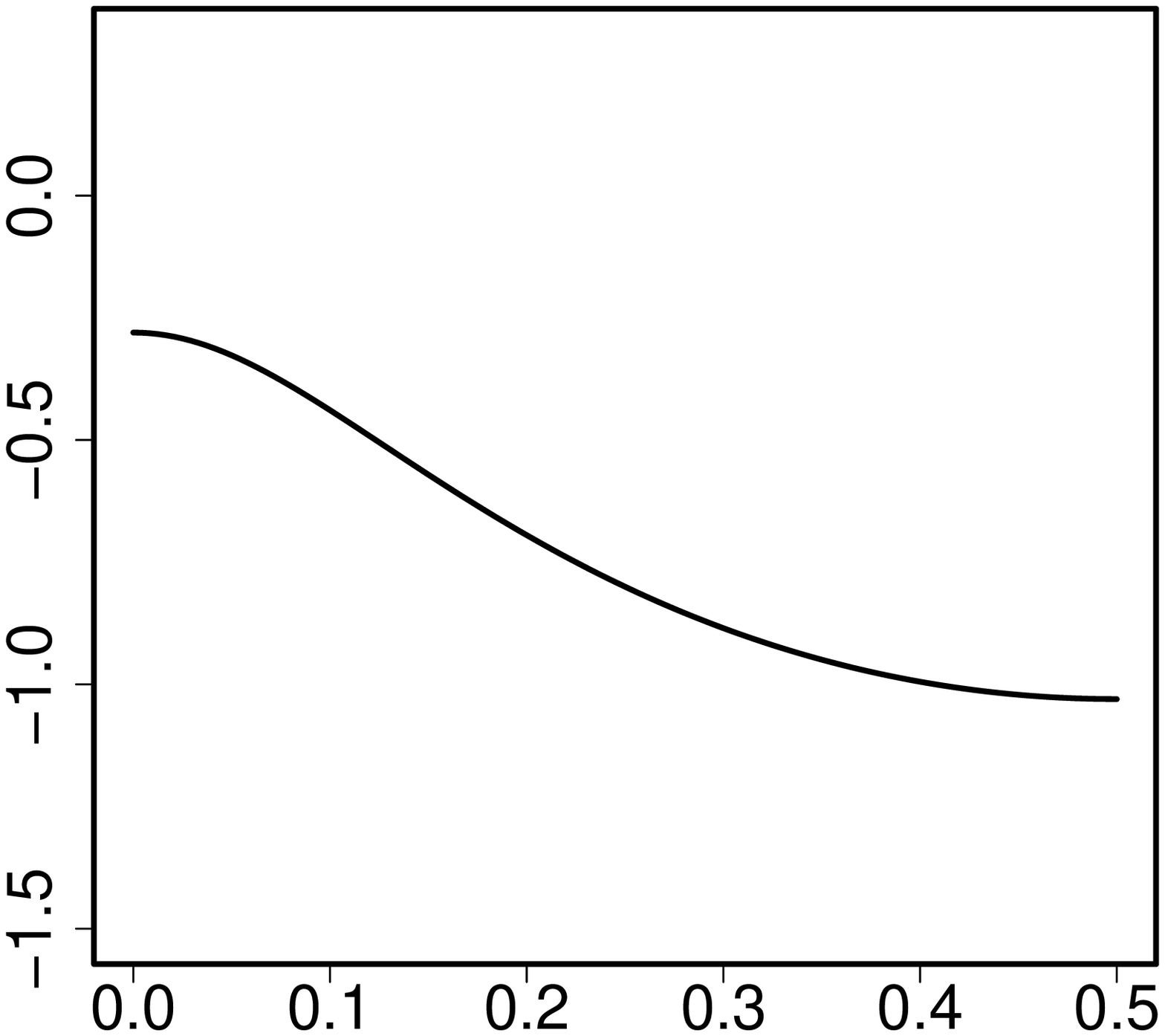}
	\rotatebox{90}{\hspace{3.6cm}\rotatebox{-90}{(1171-1270)}}\hspace{-2.4cm}
	\includegraphics[height=4cm]{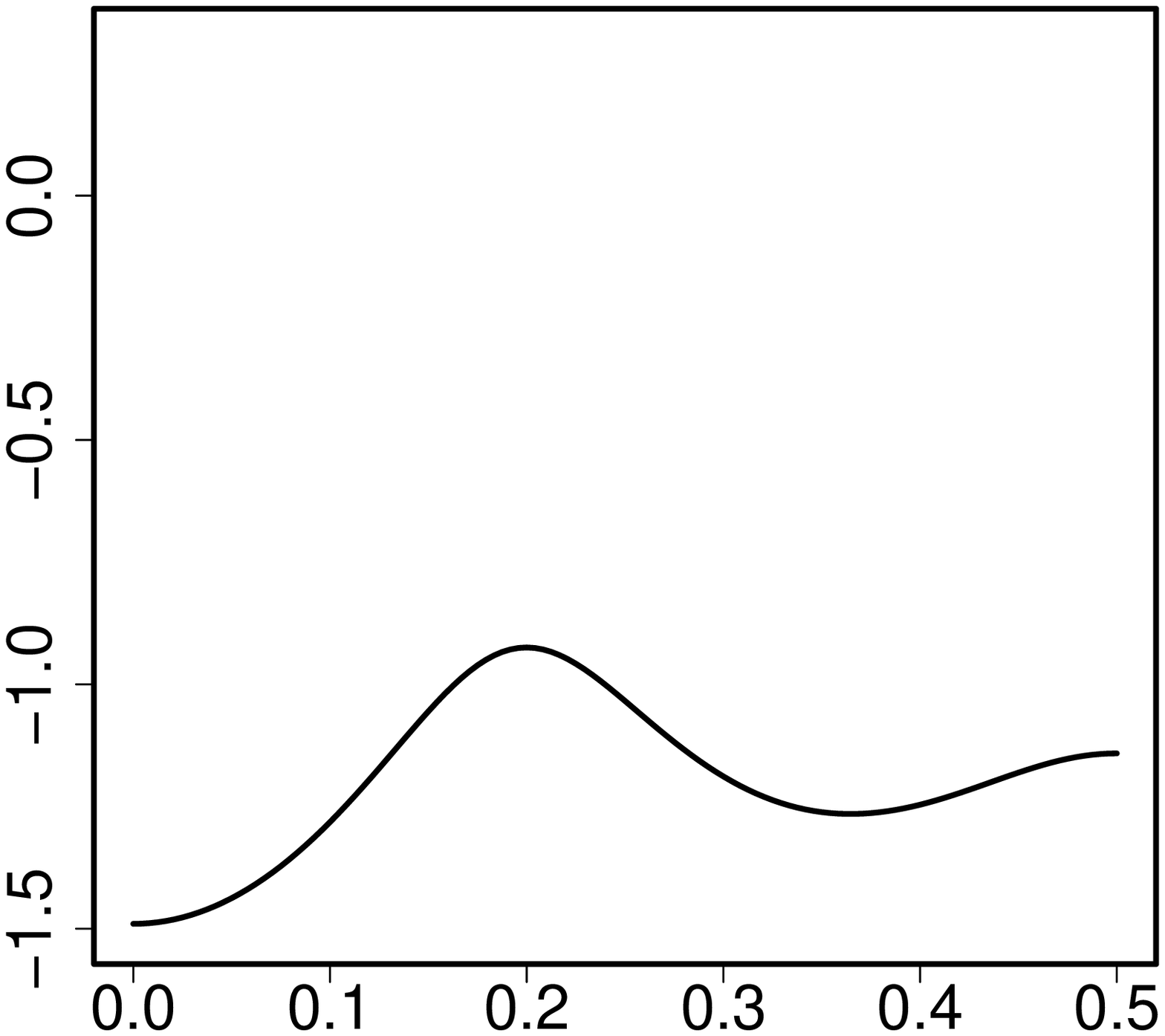}
	\rotatebox{90}{\hspace{3.6cm}\rotatebox{-90}{(1271-1370)}}\hspace{-2.4cm}
	\includegraphics[height=4cm]{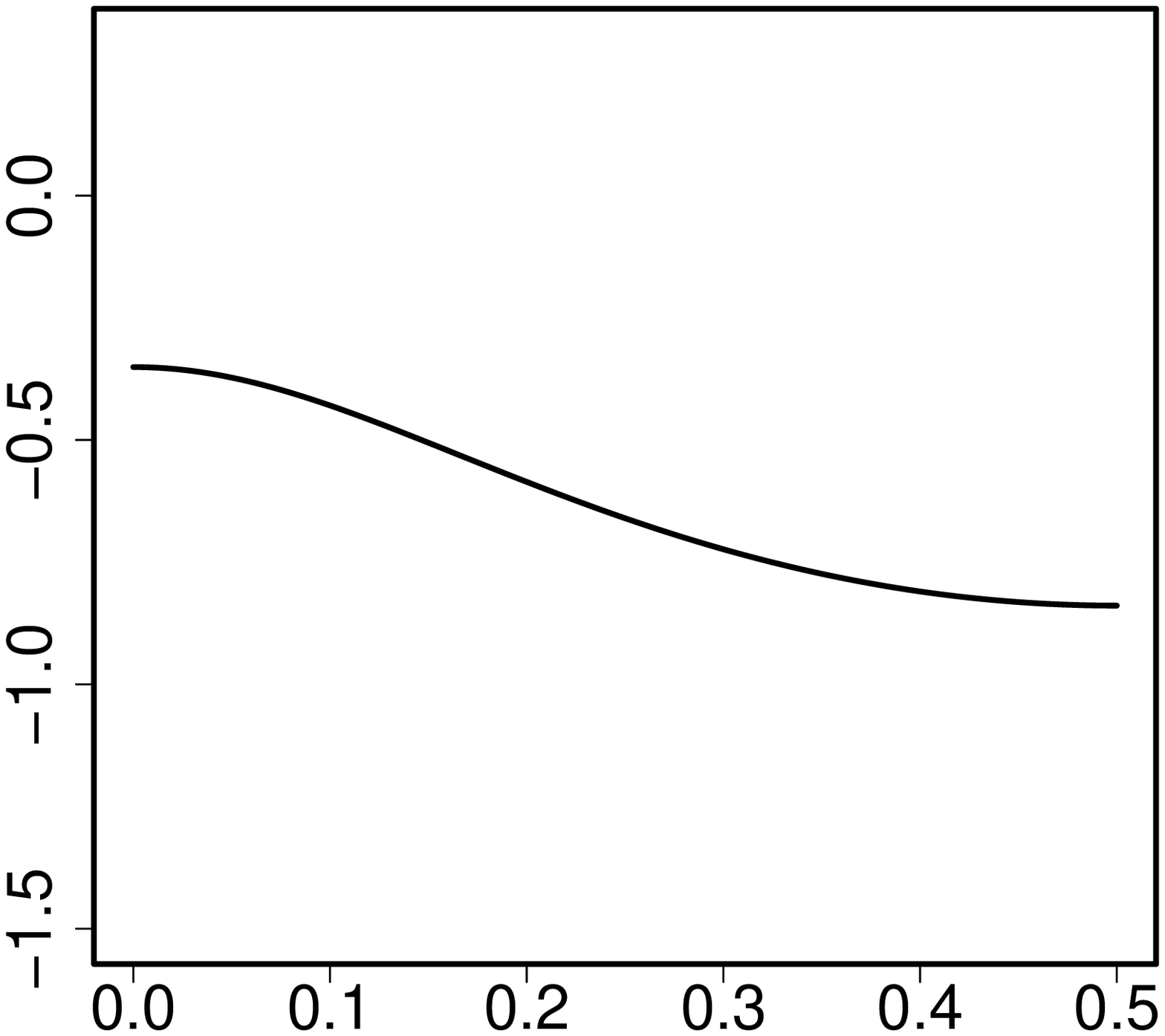}
\caption[]{Rational spectrum for the angular difference. The corresponding intervals are described at the top of each figure. Vertical axes represent $\log p(f)$.}
\label{fig.spectrum klino 100}
\end{figure}

Figure \ref{fig.spectrum klino 100} shows the spectrum of the angular difference and we show the corresponding intervals at the top of each figure. From the same reason explained above, the intervals (671-870), (871-970), (1071- 1170) and (1271-1370) imply the same styles of motion. If a fly is walking, while changing its angular component periodically and slowing the velocity component, then it explores a compact area. This happened at 171-470 steps. By changing its angular component periodically and maintaining a moderate velocity, a fly explores larger areas in the cage. This happened at 471-670 steps. If a fly walks moderately quickly while changing its angular component smoothly, then it can explore a much wider area. This happened at 671-1370 steps.

\begin{figure}[hptb]
	\hspace{3.6cm}
	\rotatebox{90}{\hspace{3.6cm}\rotatebox{-90}{(171-370)}}\hspace{-2.1cm}
	\includegraphics[height=4cm]{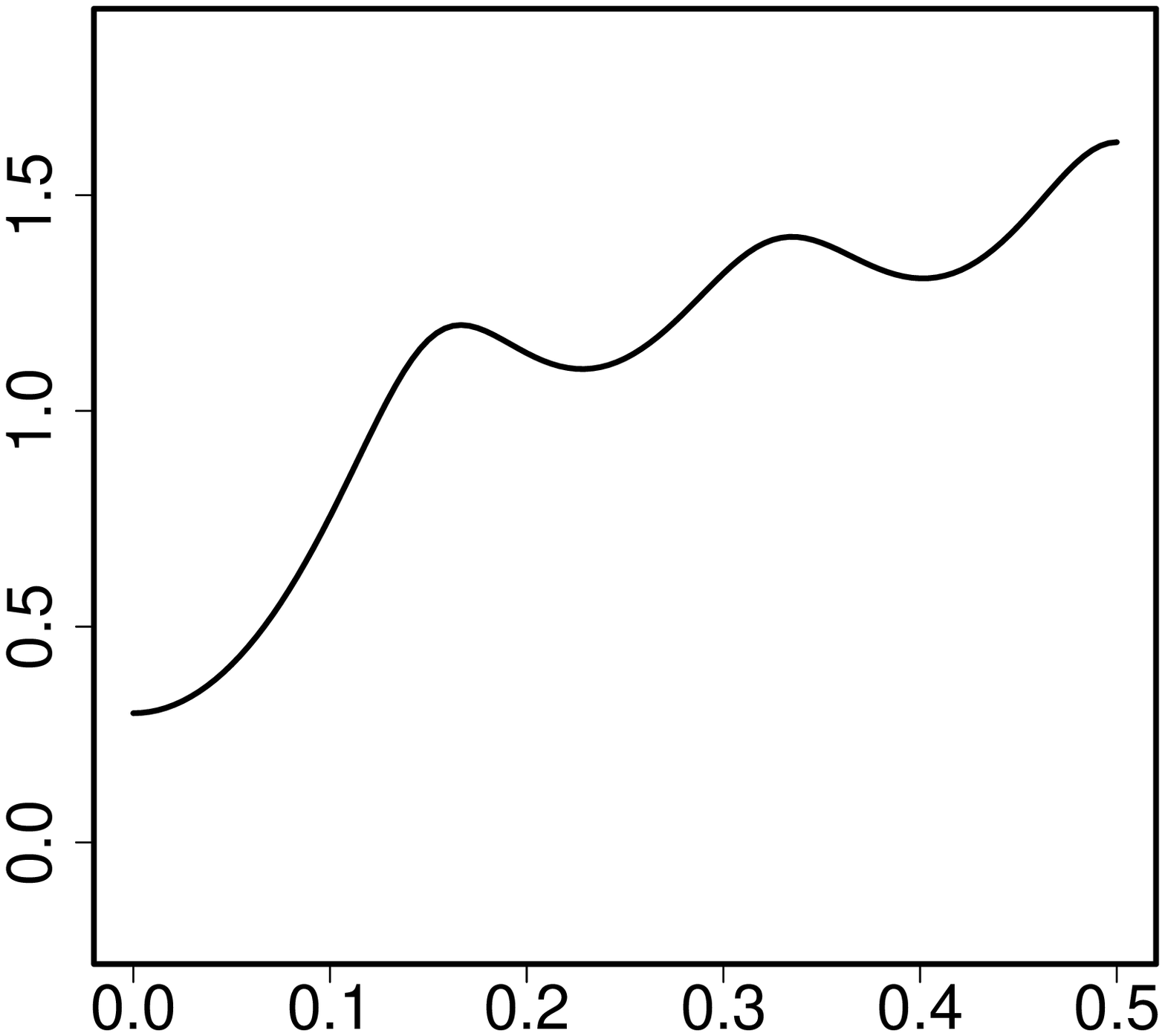}
	\rotatebox{90}{\hspace{3.6cm}\rotatebox{-90}{(371-470)}}\hspace{-2.1cm}
	\includegraphics[height=4cm]{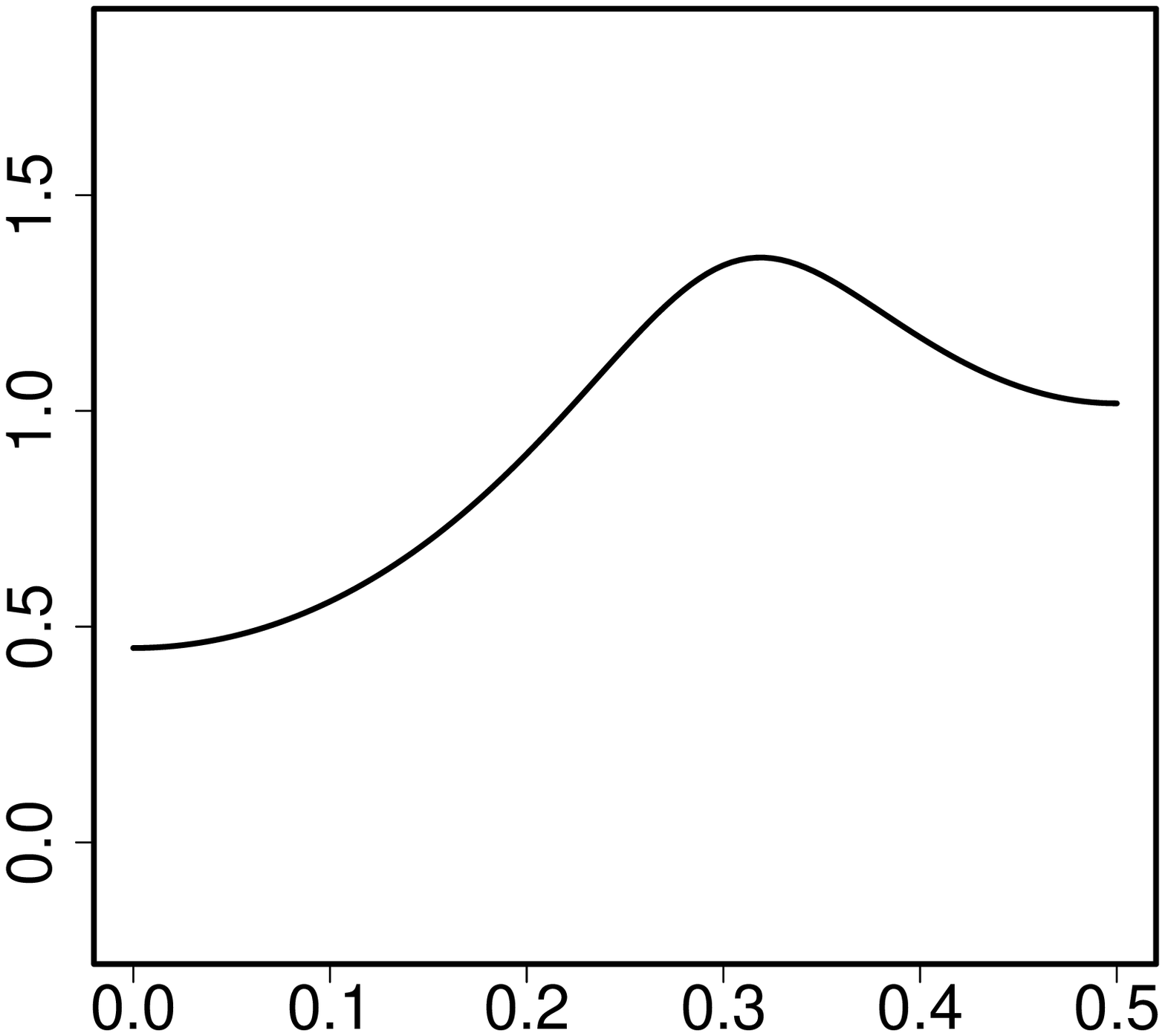}\\
	\rotatebox{90}{\hspace{3.6cm}\rotatebox{-90}{(471-570)}}\hspace{-2.1cm}
	\includegraphics[height=4cm]{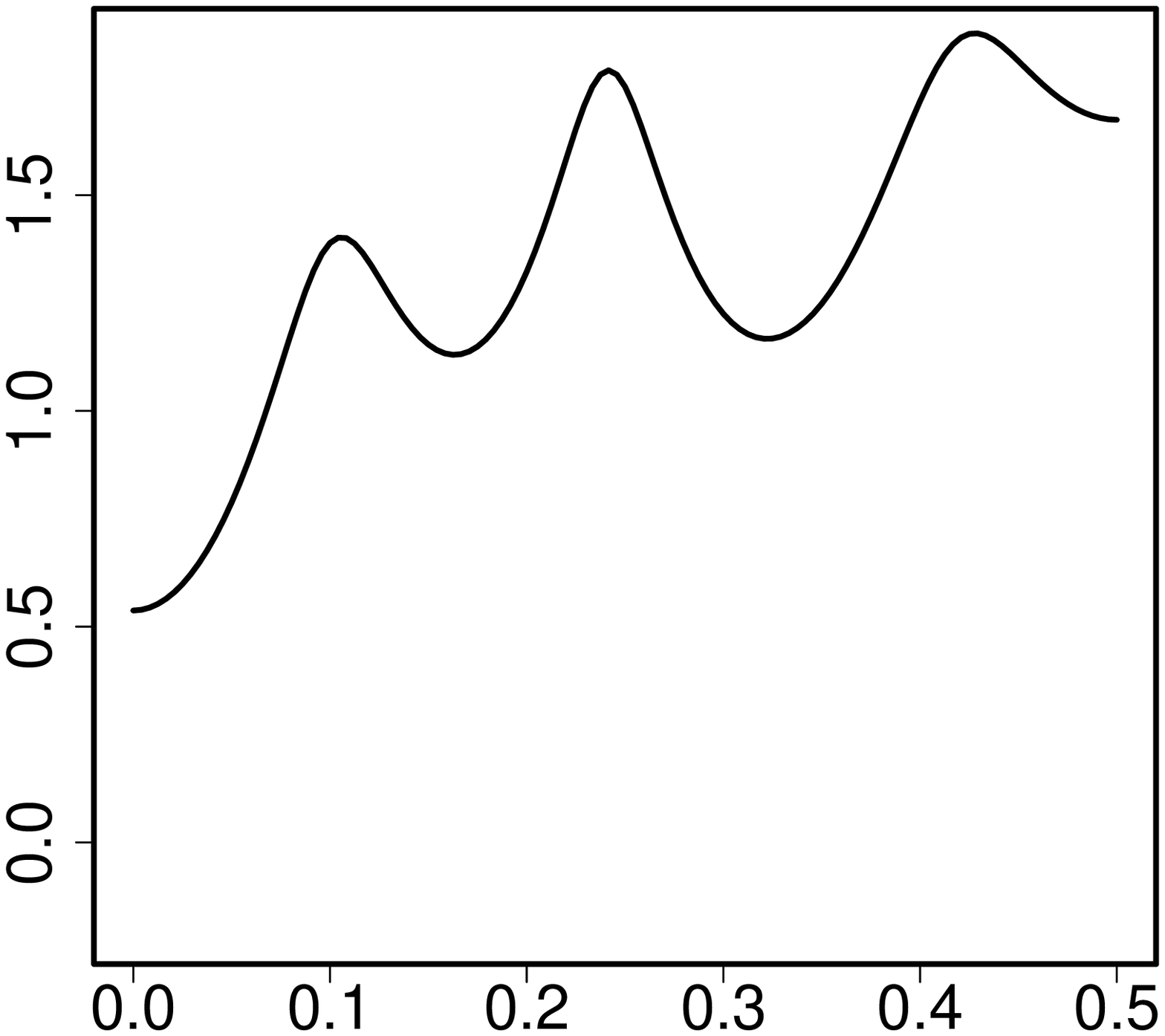}
	\rotatebox{90}{\hspace{3.6cm}\rotatebox{-90}{(571-670)}}\hspace{-2.1cm}
	\includegraphics[height=4cm]{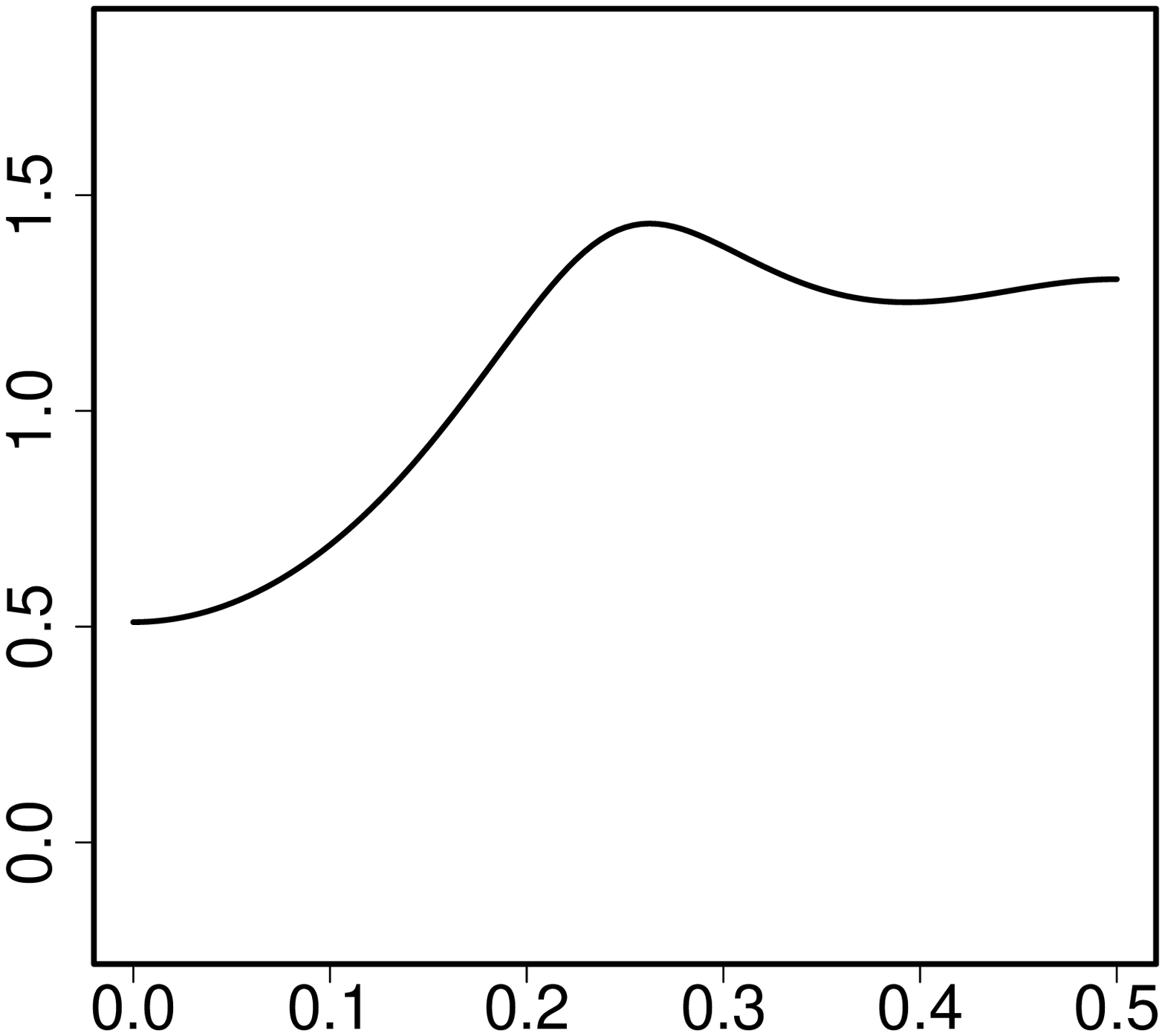}
	\rotatebox{90}{\hspace{3.6cm}\rotatebox{-90}{(671-770)}}\hspace{-2.1cm}
	\includegraphics[height=4cm]{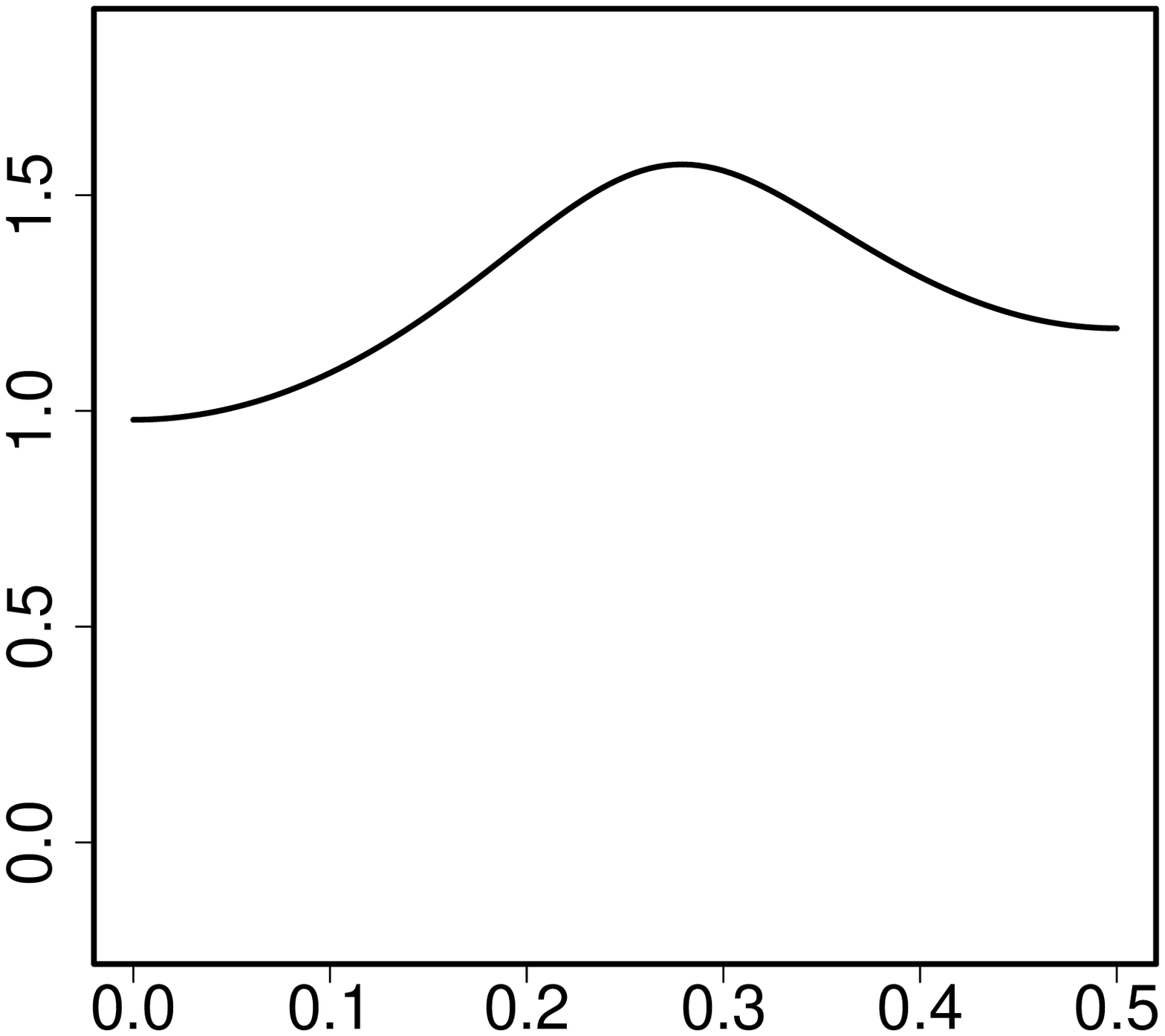}\\
	\rotatebox{90}{\hspace{3.6cm}\rotatebox{-90}{(771-870)}}\hspace{-2.1cm}
	\includegraphics[height=4cm]{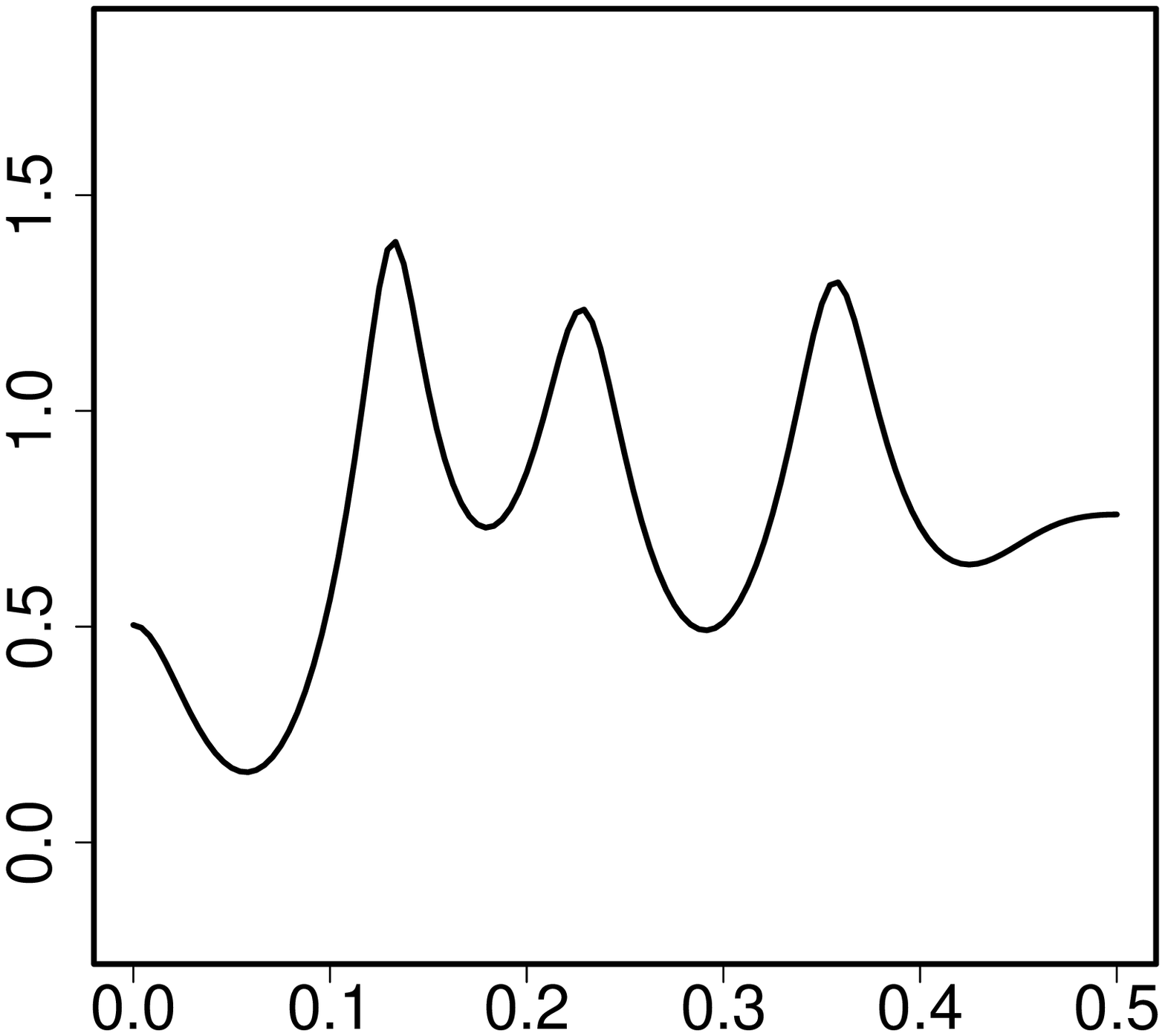}
	\hspace{3.7cm}
	\rotatebox{90}{\hspace{3.6cm}\rotatebox{-90}{(871-1070)}}\hspace{-2.4cm}
	\includegraphics[height=4cm]{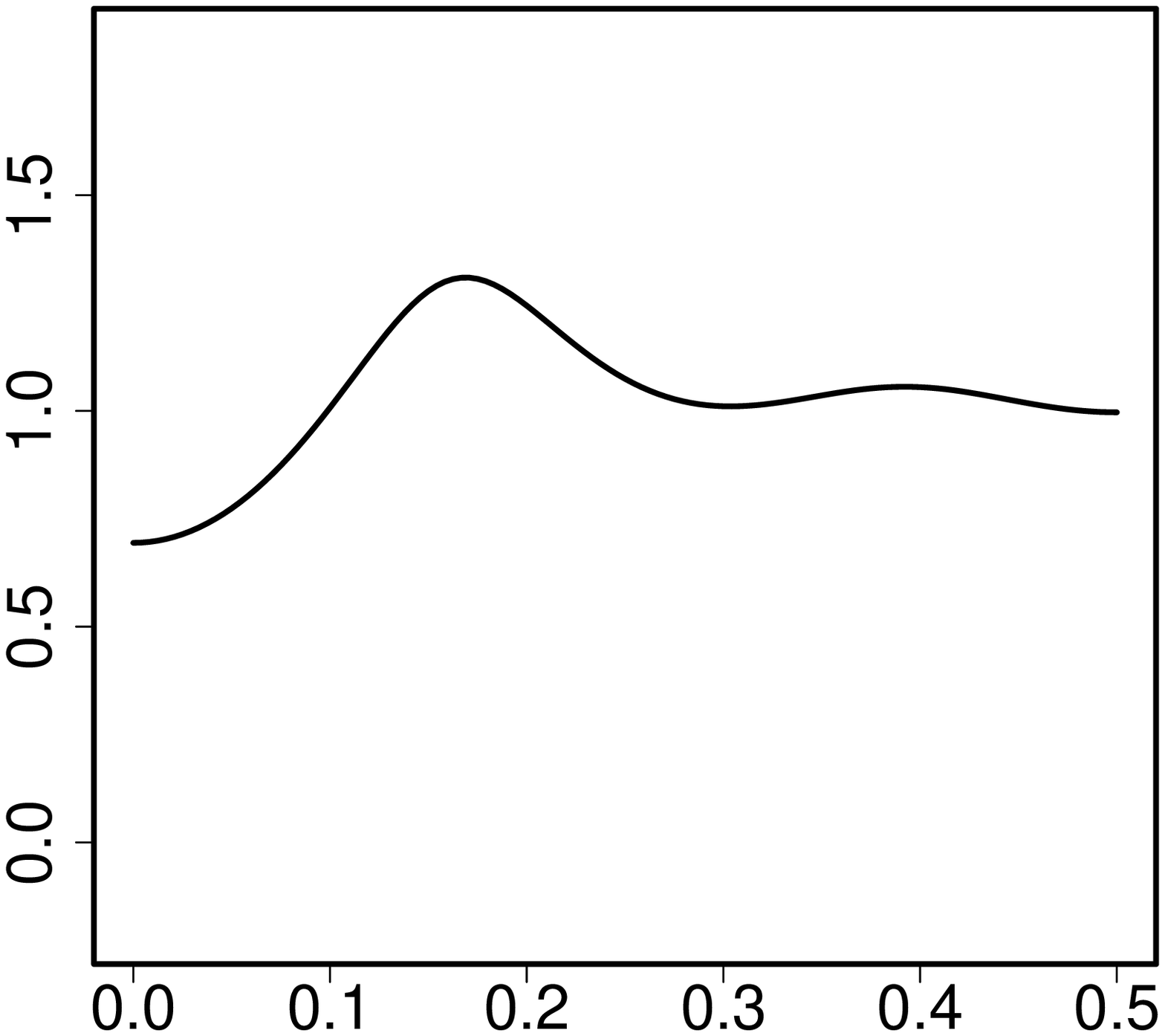}\\
	\rotatebox{90}{\hspace{3.6cm}\rotatebox{-90}{(1071-1170)}}\hspace{-2.5cm}
	\includegraphics[height=4cm]{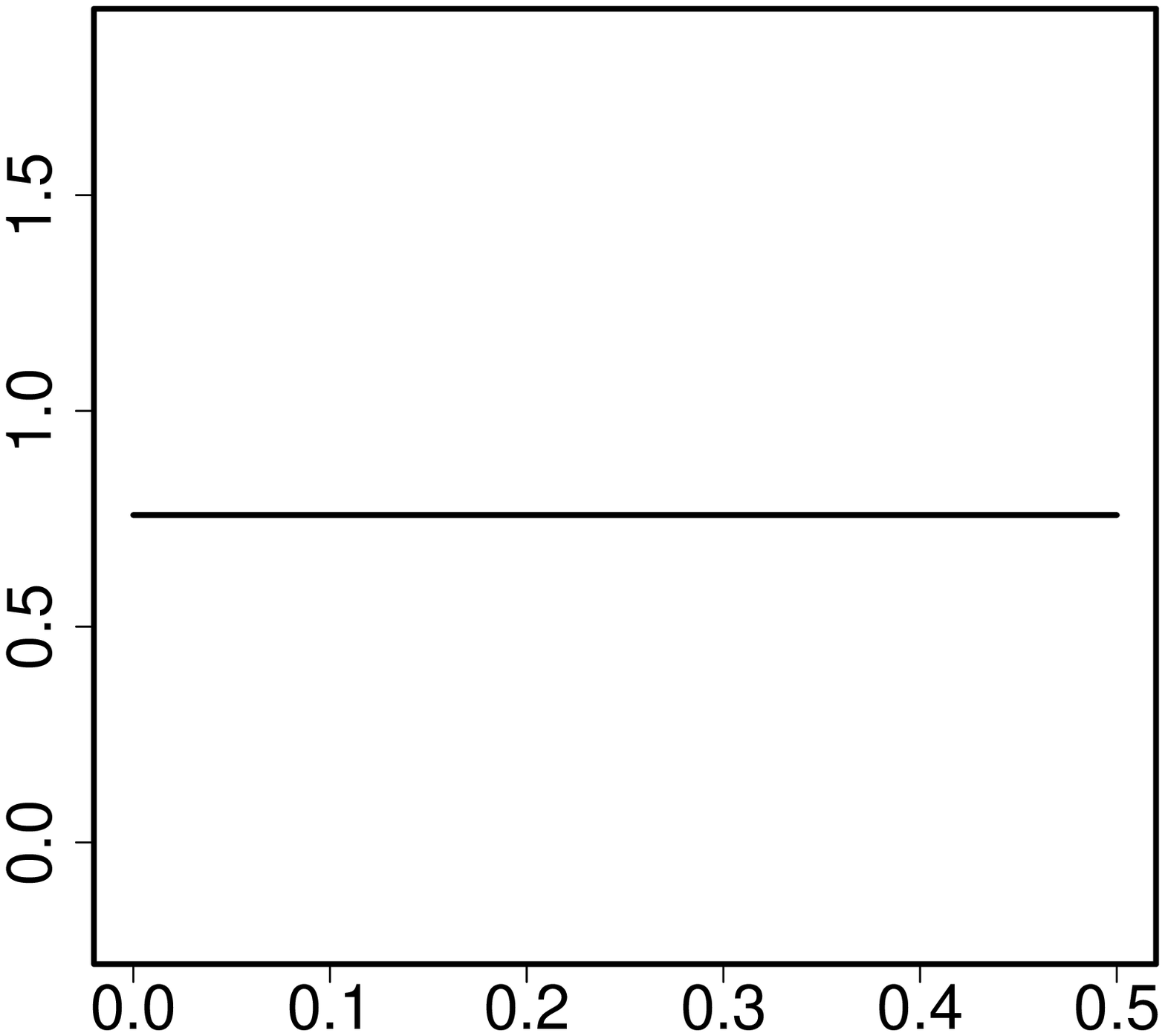}
	\rotatebox{90}{\hspace{3.6cm}\rotatebox{-90}{(1171-1270)}}\hspace{-2.5cm}
	\includegraphics[height=4cm]{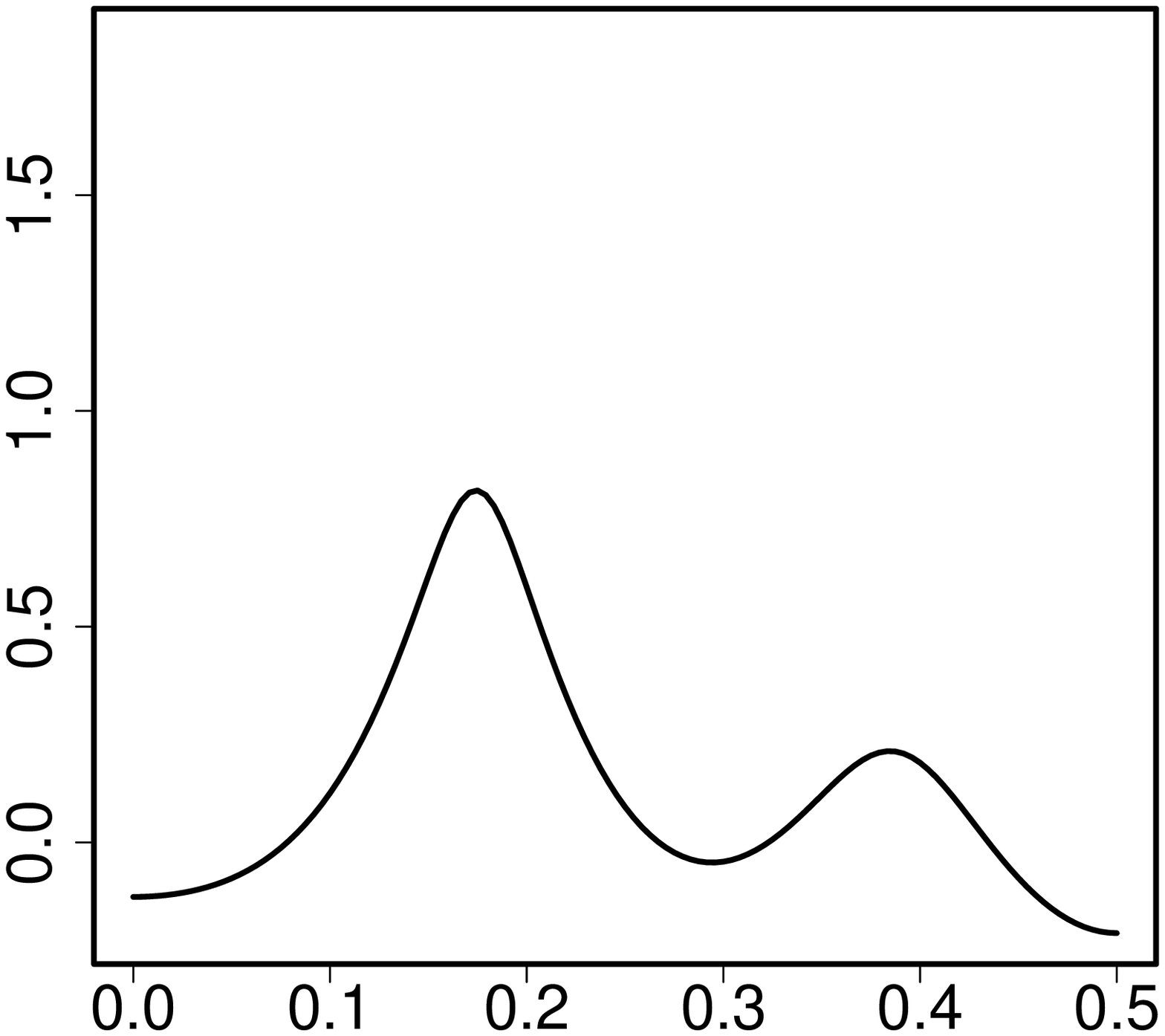}
	\rotatebox{90}{\hspace{3.6cm}\rotatebox{-90}{(1271-1370)}}\hspace{-2.5cm}
	\includegraphics[height=4cm]{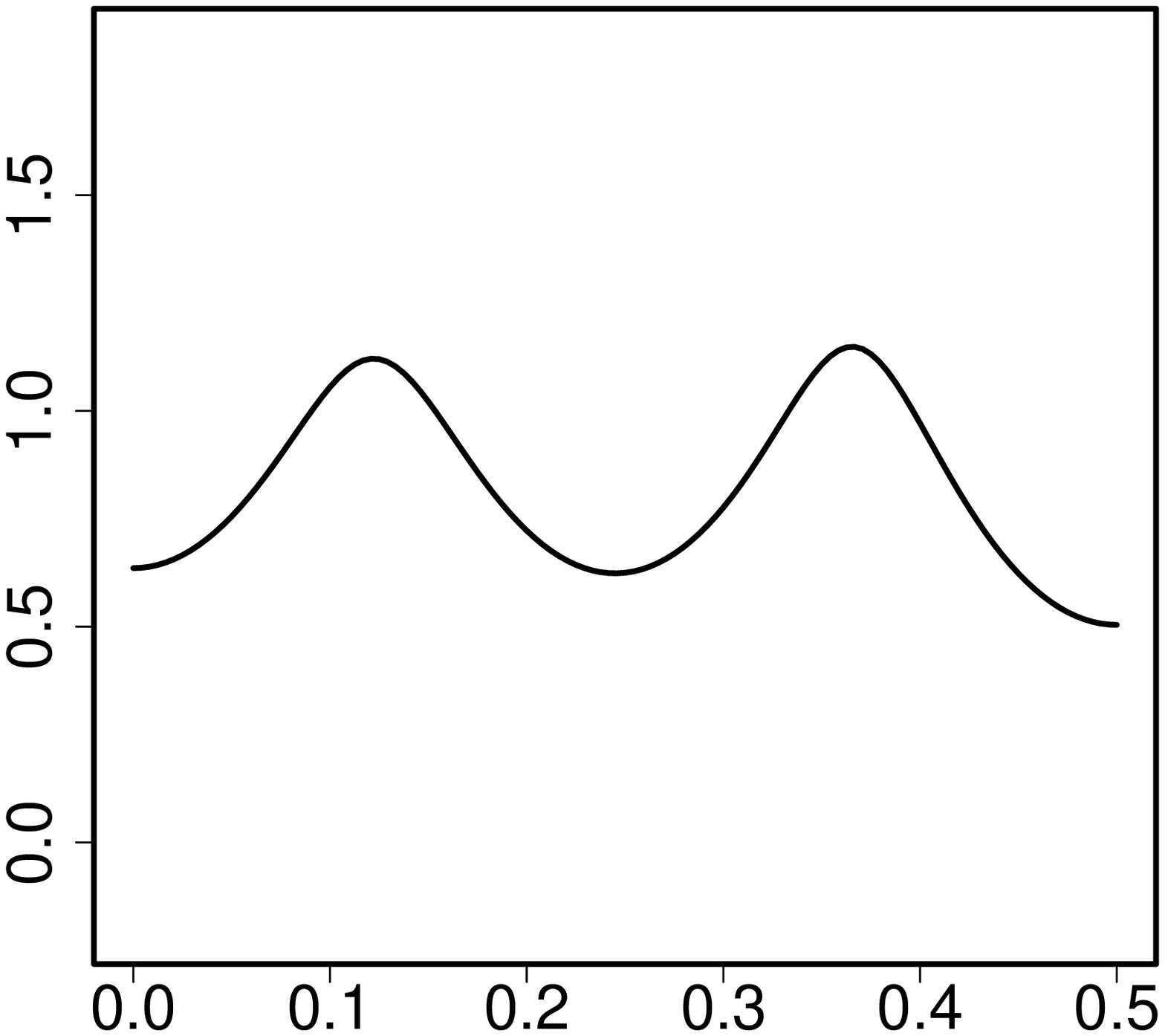}
\caption[]{Rational spectrum for the velocity's difference. The corresponding intervals are described at the top of each figure. Vertical axes represent $\log p(f)$.}
\label{fig.spectrum oltho 100}
\end{figure}

If the local AR model shows that the two time series are different from one another, this might be the result of the non-stationarity or non-linear nature of the time series, as we mentioned previously in {\S} \ref{Spectrum}.
Therefore, we should carefully estimate the point at which the fly's walking style changes using the AR spectrum.
Figure \ref{fig.spectrum oltho 100} shows the spectrum of the velocity's differences.
Finding a consistent source of the regularity apparent in these figures is difficult. We hypothesize that the figures of (371-470), (571-670) and (671-770) have similar spectrum patterns, which suggest the same walking style in those regions. Starting from this same hypothesis, we regard the intervals of (871-1070) and (1171-1270) as belonging to the same style of motion.

\subsection{Anomalous diffusion}

\begin{figure}[htbp]
\begin{center}
	\rotatebox{90}{\hspace{4cm}\rotatebox{-90}{(a)}}
	\includegraphics[height=5cm]{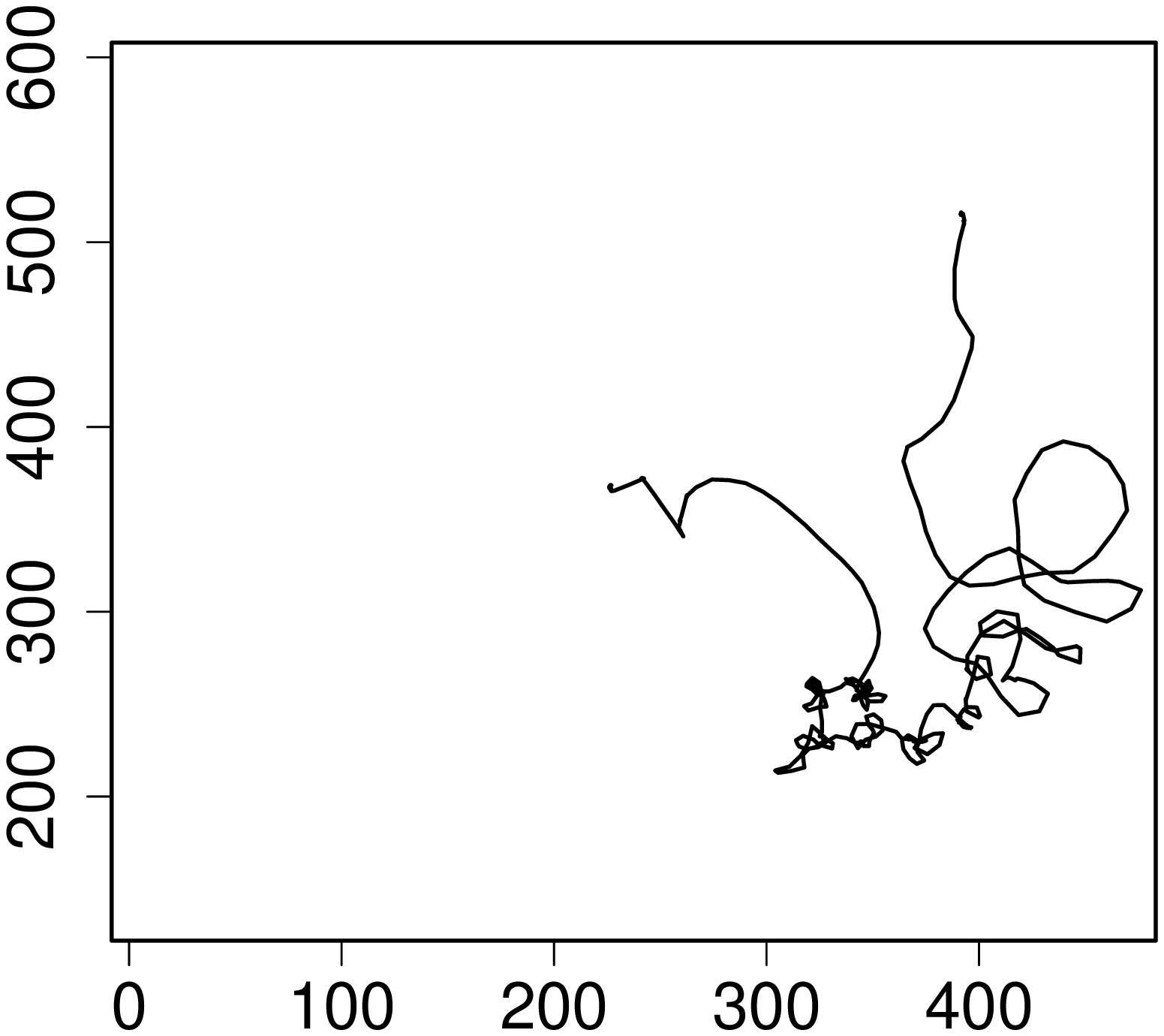}
	\rotatebox{90}{\hspace{4cm}\rotatebox{-90}{(b)}}
	\includegraphics[height=5cm]{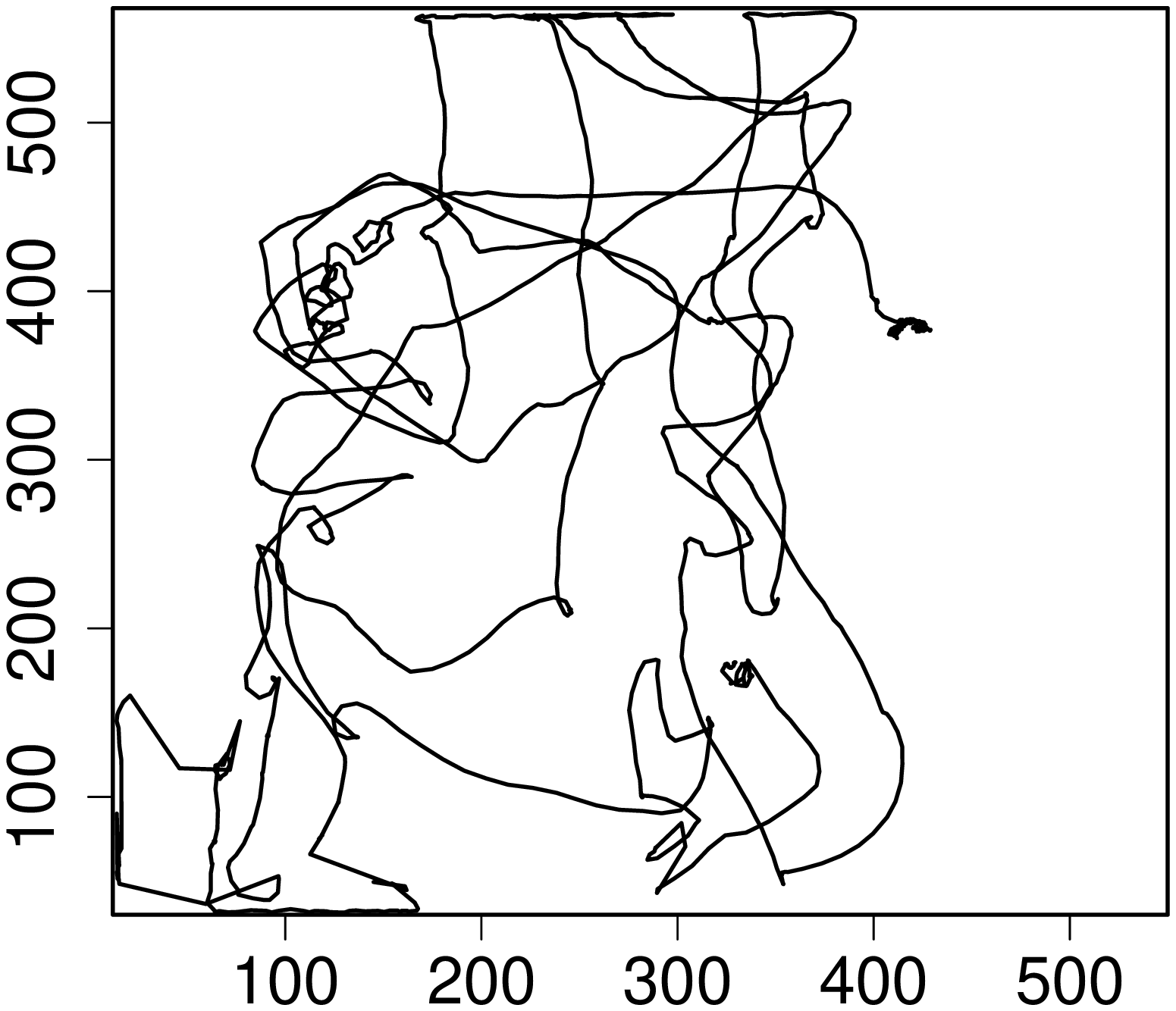}
\end{center}
\caption[]{Trajectory of a fly for about 14 minutes with sugar solution's droplets (a), and for about 27 minutes with no-sugar solution's droplets (b).}
\label{fig. trajectory food no-food}
\end{figure}

Figure \ref{fig. trajectory food no-food} shows two different time series taken from two other individuals.
Figure \ref{fig. trajectory food no-food} (a) is the trajectory for about 14 minutes with sugar droplets, i.e. the only differences from the previous series was  the positions of the droplets.
Figure \ref{fig. trajectory food no-food} (b) for about 27 minutes with no sugar  droplets.
In the case of the figure (a), the fly fed several times at around (340,250) and left from that point.
The trajectory of the series performed with the sugar solution, as shown in  Fig \ref{fig. trajectory food no-food} (a), shows some similarity to Fig. \ref{fig.trajectory 051207all}.
The fly walks around food, so that the line of its trajectory becomes crowded and zigzagged.
After the fly left from the feeding area, the line becomes smoother. In the case of Fig. \ref{fig. trajectory food no-food} (b), with a no-sugar condition, the line becomes smoother and more spread-out as in the case of Fig. \ref{fig.trajectory one minute} (b).

\begin{figure}[hptb]
	\begin{minipage}[]{5cm}
	~(a)\vspace{-2mm}\\
	\includegraphics[height=5cm]{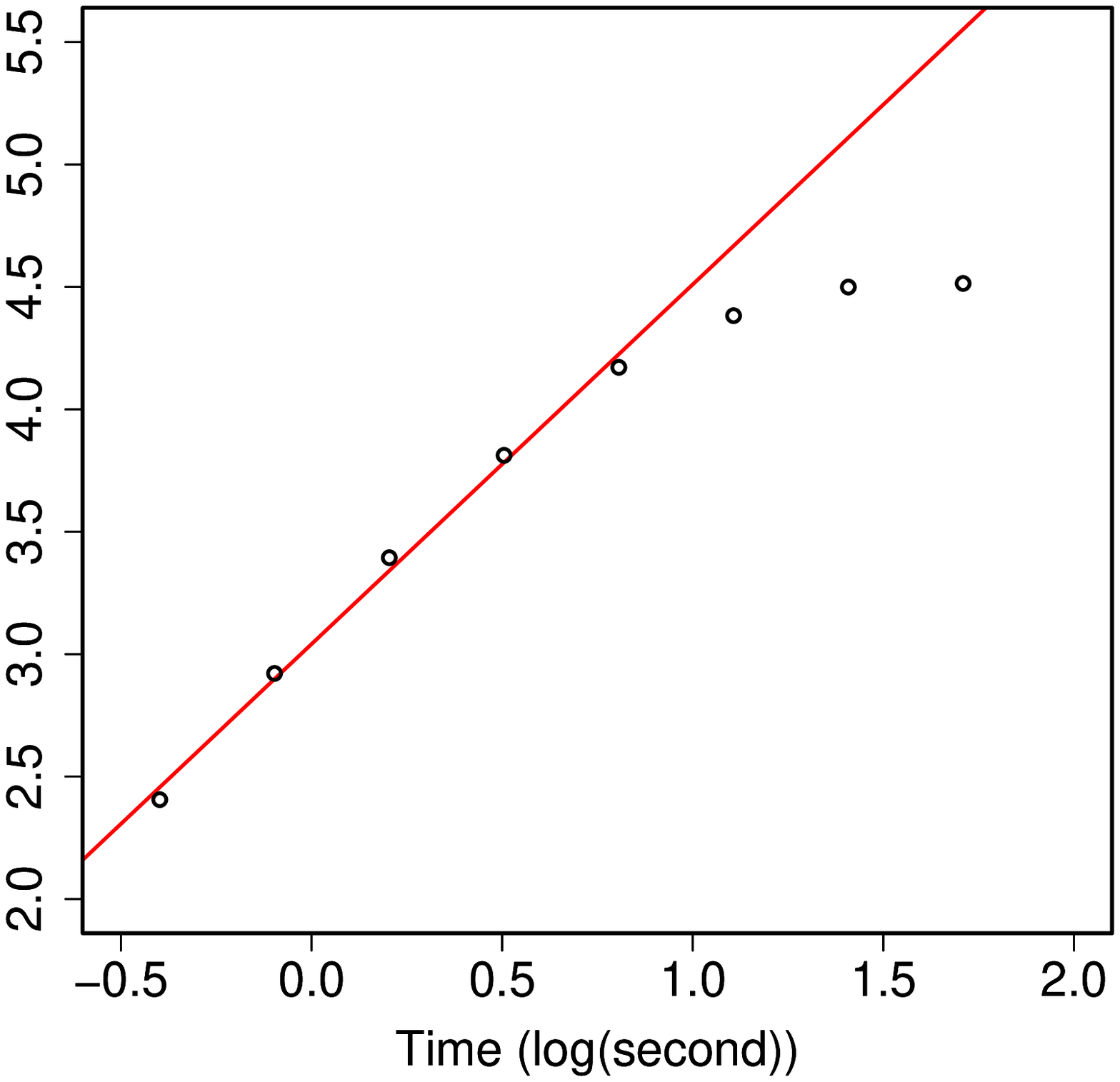}
	\end{minipage}
	\begin{minipage}[]{5cm}
	~(b)\vspace{-2mm}\\
	\includegraphics[height=5cm]{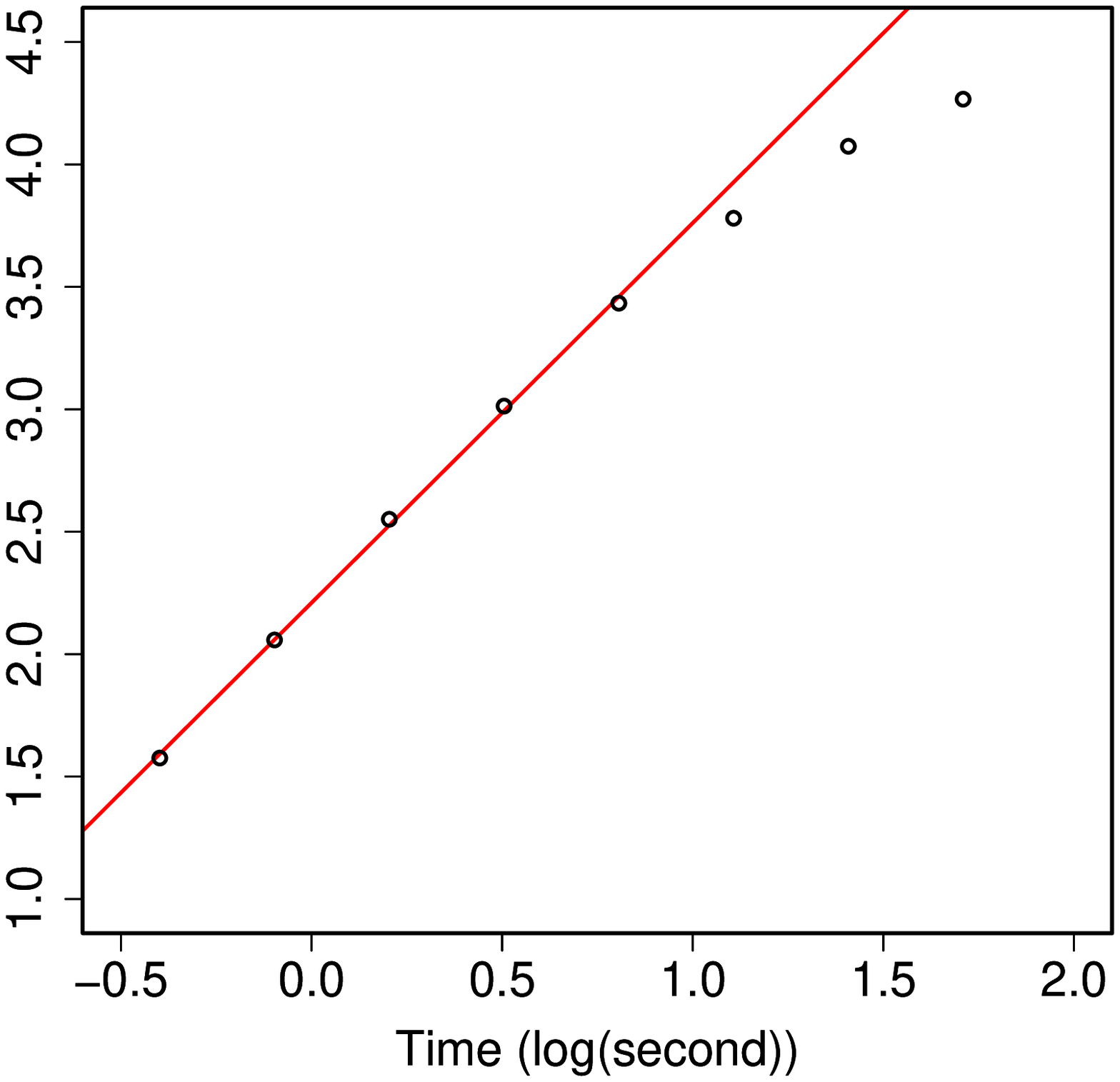}
	\end{minipage}
\caption[]{Log-log plots for the eq.~(\ref{eq. anomalous diffusion}) with the duration 0.4, 0.8, 1.6, 3.2, 6.4, 12.8, 25.6, 51.2 seconds and the line represents $ \alpha = 1.47 $ in Fig. (a) and $ \alpha = 1.55 $ in Fig. (b).
Figure (a) is corresponding the case of with sugar solution: Fig. \ref{fig.trajectory 051207all}, and Fig (b) is with no sugar solution case: Fig. \ref{fig. trajectory food no-food} (b). Vertical axes are $\log E_{t_0}[x(t+t_0)-x(t_0)]$.}
\label{fig. Anomalous diffusion density}
\end{figure}

Figure \ref{fig. Anomalous diffusion density} shows log-log plots for eq. (\ref{eq. anomalous diffusion})  and shows that the points fit well from the  time interval 0.4 sec to 8 sec. We have the exponent $\alpha = 1.47$ and $1.55$ for with and without sugar cases respectively by using these data points. This result shows that the fly's walk is indeed an anomalous diffusion. Interestingly, the navigation patterns also look different in these two cases but the anomalous diffusion appears independent of the sugar drops. 
This observation implies that the anomalous diffusion can emerge independent of the existence of food.
More likely is the possibility that the anomalous diffusion evident here is the outcome of the intrinsic randomness of a fly's motion.

In this experiment, a fly was confined in an acrylic cage, so that the anomalous diffusion is held only in a limited range.\cite{Takahashi(2003),Mantegna}  We were able to verify that the fly's walk is characterized by anomalous diffusion in the range from millimeters to centimeters. This diffusive range is at the upper limit of the range where we can verify diffusion in this experiment. This limit can be improved if we use a bigger cage, but obviously then we have to take into account a fly's ability to fly.

\section{Concluding remarks}

We found that the velocity and angular elements of a fly's motion can be independent and change asynchronously judging from the AR dimensions. Based on our observation, a fly has a control system consisting of two main choices of navigational patterns: one for changing its velocity, and another for changing its angular motion. There seem to be several channels for each choice, and the fly chooses one channel for each choice, as observed from the data set shown here. Those channels are not used in a fixed pattern, but can be dynamically varied, as the model demonstrates.\cite{ikegami}

We confirm here that a fly does not undertake long continuous walks, but rather stops frequently and sometimes stops walking and flies. These characteristics are complicating factors in this experiment. A fly appears to take flight when it cannot find anything interesting after a certain period of time. The big discrepancy between successive velocity differences is found at about 500 steps in Fig. \ref{fig.oltho klino 100-1400} (a), and is most likely is caused in such a way. After flying, the fly changes its walking style if we compare it with the style it was using before the flight. This possibility requires more analysis, however.

We only used the local stationary AR model for fixed intervals, that is L = 100.
But, the time series of a fly's walk does not change regularly at those points. One solution is to use the variable interval method, which requires an appreciable amount of CPU time.
Thus, in order to compromise, we used just the local stationary AR method with the fixed length. However, the fly's behavior is essentially nonlinear, so we should undertake a new kind of analysis to discern its complex nature.

The most interesting result for us was the fly's mode-switching behavior in the experiment. The simplest explanation may be to assume random walking with a memory effect. But, we also can interpret this behavior as a product of interplay between the internal dynamics of the fly and its externally-driven dynamics, which is studied theoretically as ``embodied chaotic itinerancy (ECI)''\cite{ikegami}. ECI explains such random traveling dynamics among different pseudo-attractors deterministically. The experiment we undertook here can be explained by the ECI approach. Asynchronous dynamics between angular and velocity's components provide the other evidence of such interplay. 

Recent experimental studies show that flies certainly have internal dynamics (e.g. van Swinderen B. et al.\cite{Greenspan (2004)}; A. Maye et al.\cite{plosone}). In Maye, A et al., the authors showed that there are deterministic chaos dynamics hidden in the intrinsic dynamics.
This observation of anomalous diffusion independent of the presence or absence of food demonstrates that there is an intrinsic randomness in the fly's walking behaviors, but it should be noted that no explicit itinerancy is observed in the no-food condition. Therefore the intrinsic randomness and the itinerant behavior are different phenomena, and we take these as the candidates for the most important elements of biological autonomy. Flies also possess some form of memory system, and genetic studies of learning 
and memory in flies are now progressing.\cite{Saitoe(2003)}

\section*{Acknowledgement}

The part of this work is supported by The 21st Century COE (Center of Excellence) program (Research Center for Integrated Science) of the Ministry of Education, Culture, Sports, Science, and Technology, Japan, and the ECAgent project, sponsored by the Future and Emerging Technologies program of the European Community (IST-1940).
We used the time series analysis and control program package (TIMSAC) for R to analyze the data.

\clearpage
\newpage

\clearpage

\clearpage

\begin{table}
(a)\\
\begin{tabular}{r|cll}
\hline\hline
steps & AR order & AR coefficient & $ \sigma^2 $ \\
\hline
\bf 171 - 370 & 5 & -0.521, -0.322, -0.287, -0.207, -0.173 & 12.563 \\
 371 - 470 & 2 & -0.420, -0.335 &  8.699 \\
 471 - 570 & 7 & -0.425, -0.176, -0.296, -0.121, -0.064, & 22.312 \\
           &   &  -0.318, -0.142     &  \\
 571 - 670 & 3 & -0.416, -0.366, -0.168 & 12.329 \\
 671 - 770 & 2 & -0.154, -0.271 & 19.372 \\
 771 - 870 & 9 & -0.149, -0.243, -0.190, -0.220, 0.015,  &  5.569 \\
           &   & 0.056, -0.054, 0.255,  0.210 &  \\
\bf 871 - 1070 & 4 & -0.049, -0.157, -0.169, -0.105 & 10.831 \\
1071 - 1170 & 0 & 0 & 5.737 \\
1171 - 1270 & 5 & 0.112, -0.230, -0.195, -0.188, 0.151 & 1.361 \\
1271 - 1370 & 4 & 0.044, -0.043, 0.057, -0.299 & 6.654 \\
\hline
\end{tabular}

(b)\\
\begin{tabular}{r|cll}
\hline\hline
steps & AR order & AR coefficient & $ \sigma^2 $ \\
\hline
 171 - 270 & 6 & 0.402, -0.069, 0.033, -0.194, -0.043, & 0.350 \\
           &   &  0.285 &  \\
 271 - 370 & 7 & 0.102, 0.184, -0.130, -0.003, 0.092,  & 0.581 \\
           &   &   -0.045, 0.284 &  \\
\bf 371 - 670 & 7 & 0.204 -0.1424, 0.060, -0.182, 0.193, & 0.390 \\
           &  &  -0.049, 0.174 &  \\
\bf 671 - 870 & 1 & 0.352 & 0.215 \\
 871 - 970 & 1 & 0.255 & 0.341 \\
 971 - 1070 & 0 & 0 & 0.163 \\
1071 - 1170 & 1 & 0.407 & 0.185 \\
1171 - 1270 & 3 & -0.056, -0.176, -0.176 & 0.064 \\
1271 - 1370 & 1 & 0.274 & 0.235 \\
\hline
\end{tabular}
\caption[]{Results of the local RA model for the velocity's difference (a), and for the angular difference (b).}
\label{table local AR oltho 07 L=100}
\end{table}

\end{document}